\documentclass{article}

\usepackage{arxiv}

\usepackage[utf8]{inputenc} 
\usepackage[T1]{fontenc}    
\usepackage{hyperref}       
\usepackage{url}            
\usepackage{booktabs}       
\usepackage{amsfonts}       
\usepackage[intlimits]{amsmath}
\usepackage{amssymb}
\usepackage{mathtools}
\usepackage{nicefrac}       
\usepackage{microtype}      
\usepackage{lipsum}		
\usepackage{graphicx}
\usepackage[numbers]{natbib}
\usepackage{natbib}
\usepackage{doi}
\usepackage{xcolor}
\usepackage{enumitem}
\usepackage{bm}
\usepackage{multirow}
\usepackage{placeins}
\usepackage{todonotes}
\usepackage[ruled, linesnumbered, nofillcomment]{algorithm2e}

\usepackage{color}

\newcommand\changes[1]{\textcolor{black}{#1}}

\title{Information Propagation and Encoding in\\ Solids: A Quantitative Approach \\Towards Mechanical Intelligence}

\date{\today}	

\author{\href{https://orcid.org/0009-0000-3325-1410}{\includegraphics[scale=0.06]{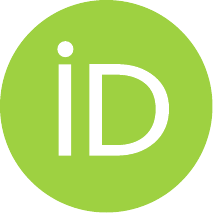}\hspace{1mm}Peerasait~Prachaseree}\\
	Department of Mechanical Engineering\\
	Boston University\\
	Boston, MA 02215 \\
	\texttt{pprachas@bu.edu} \\
	\And
	\href{https://orcid.org/0000-0001-8099-3468}{\includegraphics[scale=0.06]{orcid.pdf}\hspace{1mm}Emma~Lejeune} \\
	Department of Mechanical Engineering\\
	Boston University\\
	Boston, MA 02215\\
	\texttt{elejeune@bu.edu} \\
}

\renewcommand{\shorttitle}{Encoding Information with Mechanical System}

\hypersetup{
pdftitle={Encoding Information in Mechanical Systems},
pdfsubject={physics.data-an, cond-mat.mtrl-sci, cs.IT},
pdfauthor={Peerasait~Prachaseree,Emma~Lejeune},
pdfkeywords={mechanical intelligence, information theory, morphological computing, reservoir computing},
}

\begin{document}
\maketitle

\begin{abstract}
Engineered systems typically separate mechanical function from information processing, whereas biological systems can exploit physical structure as a medium for information processing and computation. Motivated by this contrast, recent work in mechanics has explored embedding information-processing capabilities directly into mechanical structures. However, quantitative frameworks for evaluating such capabilities remain limited. Here we address a foundational question: how does information propagate through a solid body? Using elastic bodies as a model system, we apply information-theoretic tools to treat an elastic domain as an information encoder and quantify how information transmits from applied loads to discrete sensor locations. We further connect these measures to familiar mechanical phenomena, including Saint-Venant’s effect and principal stress lines. Moving toward design, we show how geometry and architected materials can tune transmission, enabling elastic domains to either transmit or block information. Overall, this work advances quantifiable metrics and benchmark tasks for mechanical intelligence, supporting comparable designs of mechanically embodied information processing.
 
\end{abstract}
  
\section{Introduction}
\label{sec:intro}

Engineers often look to nature for inspiration in designing mechanical systems. While early efforts have focused on architecting structures to mimic the mechanical properties of natural materials \citep{wegst2015bioinspired}, more recent interest has been focused on imitating how biological structures \emph{process information}~\citep{jiao2023mechanical}. For example, soft biological tissues are fibrous networks whose microstructure gives rise not only to emergent nonlinear mechanical behaviors such as strain stiffening \citep{prachaseree2025towards, vzagar2015two}, but also to information-processing capabilities that support higher-level functions such as wound healing \citep{das2021extracellular} and cell-cell communication \citep{humphries2017mechanical, mann2019force}. This paradigm, often referred to under the umbrella terms of mechanical intelligence or mechanical computation, embeds information processing directly into the structure itself, reducing reliance on centralized computation and electronics and enabling simpler, more robust systems \citep{alu2025roadmap, kaspar2021rise, sitti2021physical}. In engineered systems, mechanically intelligent structures can improve (soft) robot functionality \citep{berrueta2024materializing} and are particularly attractive in extreme and unpredictable environments, such as within the human body as medical devices \citep{teixeira2025stimuli} or in outer space \citep{chen2019autonomous}. Motivated by these advantages, a broad range of approaches have been explored to implement mechanical intelligence, including physical reservoir computing \citep{bhovad2021physical, nakajima2020physical}, morphological computation \citep{hauser2011towards, pfeifer2009morphological}, mechanical neural networks \citep{lee2022mechanical, oktay2023neuromechanical}, locality sensitive hashing \citep{lejeune2023locality}, and mechanical logic gates \citep{meng2021bistability, parsa2022evolution}, as well as task-specific designs for grasping \citep{yang2021grasping}, locomotion \citep{liu2025dynamic}, mechanical cloaking \citep{senhora2025unbiased, wang2022mechanical}, and maze solving \citep{xi2024emergent, zhao2023physically}. While these approaches demonstrate compelling functionality, they span a wide range of problem-specific formulations and objectives, making it difficult to compare systems or assess their relative capabilities under a common notion of mechanical intelligence.

Despite the broad potential of ``mechanical intelligence'' and ``mechanical computing,'' there is currently no clear consensus on the specifics of what constitutes as mechanical intelligence. At a high level, it is generally agreed upon that for a system to exhibit intelligence, it is necessary for the system to 1) sense and perceive its surroundings, 2) retain memory about past states, and 3) apply this knowledge to react and adapt to its environment \citep{alu2025roadmap, kaspar2021rise, sitti2021physical}. In this work, as a step toward mechanical intelligence, we focus on sensing and perception of surroundings by investigating how mechanics can play a role. A central challenge in this effort is the lack of a quantitative, task-agnostic evaluation framework, which makes it difficult to compare different systems or to identify meaningful quantities to measure or optimize. The importance of a such a framework can be seen in the field of machine learning. Although learning itself is an abstract concept, machine learning has benefited from foundational theories such as Probably Approximately Correct (PAC) learning \citep{valiant1984theory}. PAC learning plays a central role in formalizing what it means for a system to learn by providing evaluation criteria that are independent of specific tasks, algorithms, or model architectures. In contrast, comparable unifying frameworks are still lacking for mechanical intelligence. With this goal in mind, we seek an evaluation criterion for mechanical sensing that is defined through the information shared between a system's input and output, rather than through any task-specific objective or loss function. We use the term \emph{task-agnostic} to demonstrate that even though the framework is currently applied to analyze sensing in static, linear elastic systems, the underlying criterion does not rely on any specific task or objective.

To realize this criterion, we turn to information theory, the mathematical study of communication and message transmission \citep{shannon1948amathematical, shannon1998themathematical}, which formalizes abstract concepts such as information and uncertainty into well-defined, measurable quantities, such as information entropy. While originally developed for communication systems, information theory has since found applications in a variety of different disciplines\citep{cover1999elements}, such as inverse problems \citep{alberts2023physics, ensslin2019information}. The use of information theory in physical sciences, however, is still limited. One of the earliest applications of information theory to physics arises in statistical mechanics, where thought experiments such as Maxwell's demon \citep{maxwell2012theory} and Landauer's Principle \citep{landauer1991information} establish fundamental links between energy and information. More recently, these principles have been extended to information engines, or active systems that convert information into work \citep{saha2021maximizing, vansaders2023informational}. In fluid mechanics, information theory have been used to model and control turbulent flows \citep{lozano2022information}. For solid mechanics, information theory has previously been used to calibrate constitutive models~\citep{bhattacharya2026optimal,ihuaenyi2025mechanics}, and analyze strain energy distributions in mechanical metamaterials \citep{klein2021spectral}. Despite high promise, information theory has not yet gained a footing in the study of mechanical intelligence. Motivated by the need for a quantitative, task-agnostic evaluation, we develop an information-theoretic framework that characterizes how mechanical structures encode and transmit information. Since this framework is new to mechanics, we begin with linear elastic systems as a tractable first setting. We envision this work as a foundation on which more complex studies of mechanical intelligence can be built.

The remainder of the paper is organized as follows. We begin by introducing key concepts from information theory and formulate information propagation in elastic solids as an information channel, with the applied load serving as the input signal (Section~\ref{sec:meth}). We then physically validate this mechanical information channel using an elastic halfspace problem in Section~\ref{result:halfspace}, and connect information propagation to Saint-Venant’s effect. We then consider architected materials, where Section \ref{result:architected} examines how modifications to the domain geometry affects information propagation in elastic solids. Next, in Section \ref{result:optimize}, we demonstrate that the proposed framework can be used to optimize for structures that maximize and minimize information flow. \changes{Finally, we demonstrate a preliminary study on extending our frame to nonlinear mechanics in Section \ref{result:elastica}.} We end by presenting our conclusions and future outlook in Section \ref{sec:conclusion}.

\section{Methods}
\label{sec:meth}

While there have been efforts to embed information processing capabilities in mechanical systems \citep{sitti2021physical, yasuda2021mechanical}, to our knowledge, there are no clear established \emph{quantitative metrics} to measure information processing in these systems. In fact, the even more fundamental question of how information about an applied load propagates through an elastic solid remains open. This work establishes a unified, quantitative framework for information propagation in mechanical systems, laying the groundwork for future studies. To this end, we first provide a brief introduction to key concepts from information theory in Section \ref{meth:info_theory}, with a more comprehensive discussion given in Appendix \ref{appendix:info_theory}. Building on these information-theoretic definitions, in Section \ref{meth:info_channel}, we introduce the information channel, a formal framework for studying information transmission with tools from information theory. We then present our strategy for leveraging the information channel to quantify how elastic bodies propagate information about the applied load.

\subsection{Mutual Information as a Measure of Information Transmission}
\label{meth:info_theory}
Information theory is a unified mathematical framework to study and quantify information during transmission and storage \citep{cover1999elements, shannon1948amathematical}. While initially formulated in terms of discrete random variables (see Appendix \ref{appendix:dis_info}), information theory has also been applied to continuous systems \citep{shannon1998themathematical}.  The core measure in information theory is information entropy, or differential entropy in the continuous case, which characterizes the amount of information needed to describe the state of the random variable in question. A random variable with larger information entropy means that more information is needed to fully represent its current state, which can also be interpreted as the random variable having a higher amount of uncertainty. On the other hand, a random variable with zero entropy is deterministic \footnote{Note this condition is true only if the random variable has finite precision}. More formally, for a continuous random variable $\bm{X} \in \mathbb{R}^{d_x}$ with the corresponding probability density function $f(\bm{x})$, the differential entropy of $\bm{X}$ is defined as:

\begin{equation}
h(\bm{X})  = -\int_\mathcal{X} f(\bm{x})\log \frac{f(\bm{x})}{m(\bm{x})} \, dx \, ,
\end{equation}

where $\mathcal{X}$ is the support set and $m(\bm{x})$ is the ``invariant measure'' to ensure coordinate independence and dimensional consistency. Note that in this work we use the natural logarithm to compute information quantities, although other logarithmic bases are also acceptable in general. More information on the derivations, necessity, and interpretations of $m(\bm{x})$ is found in the Appendix \ref{appendix:meth_diffentropy}. Subsequently, in the case of multiple variables, the joint and conditional relative entropies are defined as follows:

\begin{subequations}
\begin{align}
h(\bm{X},\bm{Y}) &= -\int_\mathcal{Y} \int_\mathcal{X} f(\bm{x},\bm{y})\log \frac{f(\bm{x},\bm{y})}{m(\bm{x},\bm{y})} \, dx \, dy \\
h(\bm{X}|\bm{Y}) &= -\int_{\mathcal{Y}}\int_{\mathcal{X}} f(\bm{x},\bm{y}) \log \frac{f(\bm{x}|\bm{y})}{m(\bm{x}|\bm{y})} \,dx \,dy \, ,
\end{align}
\end{subequations}
where $\bm{Y} \in \mathbb{R}^{d_y}$ is the additional random variable, $f(\bm{x},\bm{y})$ denotes the joint continuous probability density function, $f(\bm{x}|\bm{y})$ denotes the continuous conditional probability density function, and $\mathcal{Y}$ denotes the support set of $\bm{Y}$. Intuitively, the joint entropy $h(\bm{X},\bm{Y})$ measures the total uncertainty of $\bm{X}$ and $\bm{Y}$ together, while the conditional entropy $h(\bm{X}|\bm{Y})$ measures the remaining uncertainty on $\bm{X}$ after observing $\bm{Y}$. However, while differential entropy is useful for quantifying uncertainty in random variables, mutual information is used to quantify \emph{relationships} between random variables. In terms of entropy, mutual information can be defined as:

\begin{subequations}
\label{eqn:MI_ent}
\begin{align}
I(\bm{X};\bm{Y}) &= h(\bm{X}) + h(\bm{Y}) - h(\bm{X},\bm{Y}) \label{eqn:MI_entropyfull}\\
&= h(\bm{X})-h(\bm{X}|\bm{Y}) \label{eqn:MI_entX}\\
&= h(\bm{Y}) - h(\bm{Y}|\bm{X}) \label{eqn:MI_entY} \, .
\end{align}
\end{subequations}

Here, Eqn.\eqref{eqn:MI_entX} gives an intuitive interpretation of MI: the reduction of entropy (i.e., uncertainty) in a random variable $\bm{X}$ due to observing of $\bm{Y}$. Due to the symmetric nature of MI, Eqn. \eqref{eqn:MI_entY} also describes the reduction of uncertainty in $\bm{Y}$ after observing $\bm{X}$. As a result, MI is often referred to as \emph{information gain}. In this work, we use MI to quantify the efficacy of information transmission in elastic bodies. Additional details of our strategy to estimate entropy and mutual information are presented in Appendix \ref{appendix:ksg}.

\subsection{Information Channels in Mechanics}
\label{meth:info_channel}

\begin{figure}[t]
    \centering
    \includegraphics[width= 0.5\textwidth]{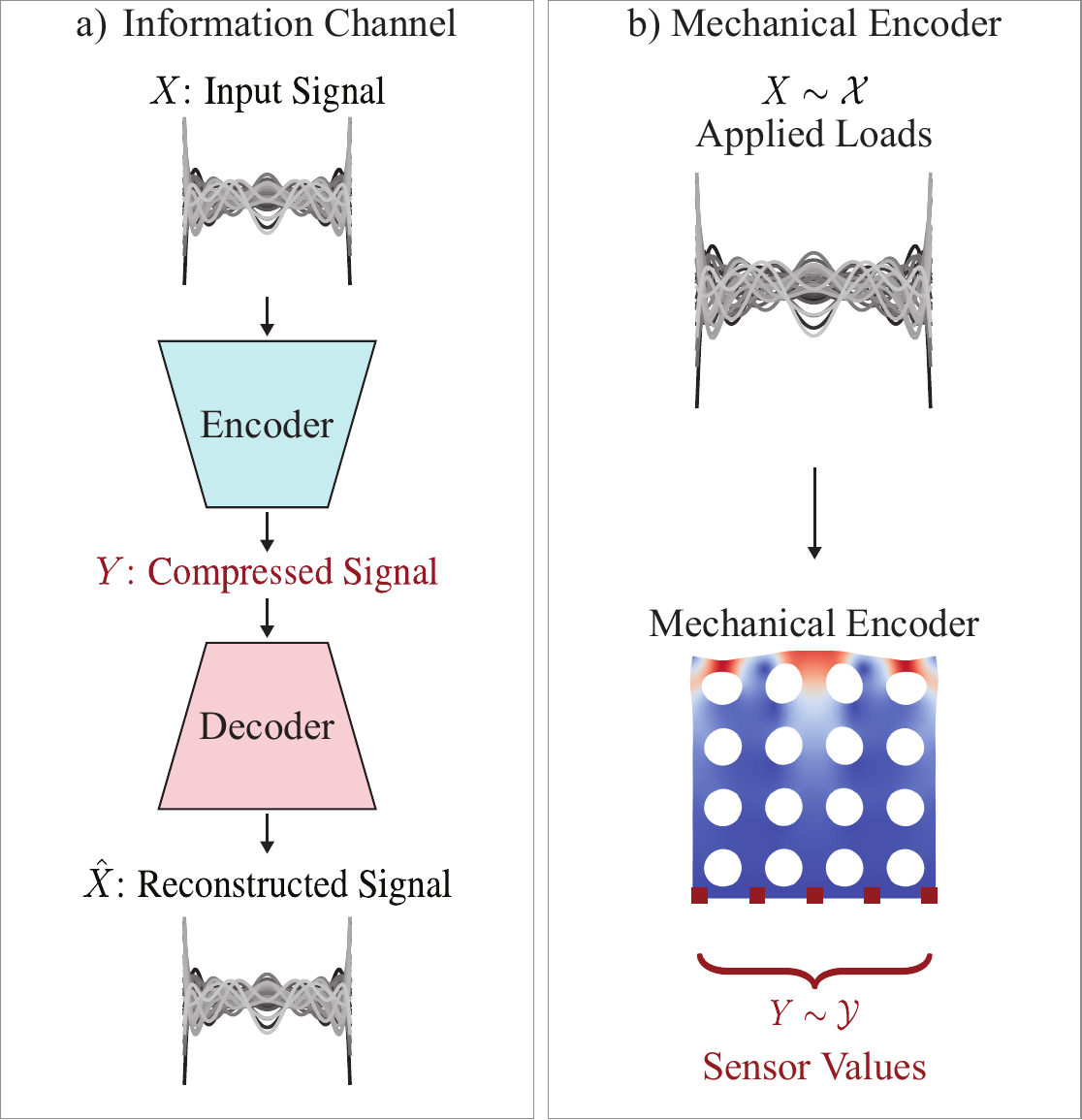}
    \caption{Interpretation of a mechanical system as an information channel. a) Block diagram of a simple information channel. Input signal $\bm{X} \sim \mathcal{X}$ is compressed to signal $\bm{Y} \sim \mathcal{Y}$ through an encoder. $\bm{Y}$ is then transmitted and passed through a decoder to obtain the final decoded signal $\hat{\bm{X}} \sim \hat{\mathcal{X}}$. b) Schematic of an elastic solid as an information encoder. The load $\bm{X} \sim \mathcal{X}$ is the input signal and the elastic solid body acts as an encoder. The reaction forces measured at discrete sensor points are the output compressed signal $\bm{Y} \sim \mathcal{Y}$. }
    \label{fig:meth_info}
\end{figure}

With the information measures defined, we are now ready to introduce the main scientific question of this work: can we use an elastic solid to encode and transmit information? Toward answering this question, we define a simple information channel and extend the definition to a mechanical system that uses a solid elastic body as an information encoder. 
An information channel is a formal mathematical model used to analyze information transmission between input signals and output signals.
To start, we define a simple noiseless communication channel that consists of:

\begin{enumerate}
\item{An input source signal $\bm{X}$ that is modeled as a random process sampled from space $\mathcal{X}$.}
\item{An encoder that transforms the input source signal $\bm{X}$ into a compressed signal $\bm{Y}$ that is suitable to be transmitted. The space of possible values of $\bm{Y}$ is denoted as $\mathcal{Y}$.}
\item{A decoder operating on the compressed message $\bm{Y}$ to obtain a reconstructed message $\hat{X}$ which is an approximation of the true message $\bm{X}$. The space of possible $\hat{\bm{X}}$ values is denoted as $\hat{\mathcal{X}}$. }
\end{enumerate}

This notation is schematically illustrated in Fig. \ref{fig:meth_info}a.
As a start towards understanding mechanical information processing, we focus on the encoder portion of the information channel and investigate the efficacy of a solid elastic body as an encoder since this is the first component responsible for transmitting information. In this case, the input source $\bm{X} \sim \mathcal{X}$ is the applied load and the output measured reaction force is $\bm{Y} \sim \mathcal{Y}$ (see Fig. \ref{fig:meth_info}b).
To mimic physical systems, we will measure reaction forces at discrete sensor points in the solid body such that $k$ sensor readings can be represented in $d_y$-dimensional vector $\bm{Y} \in \mathbb{R}^{d_y}$ (see Fig. \ref{fig:meth_info}b). For simplicity, these sensor readings are noiseless and can record reaction forces at floating-point precision. Extensions to noisy systems can be explored in future work.

Since the sensors are noiseless, we are able to infer some information-theoretic bounds for our mechanical information encoder. For an arbitrary set of sensors $\bm{Y}$, the conditional entropy $h(\bm{Y}|\bm{X})=0$ because $\bm{Y}$ can be inferred exactly from $\bm{X}$. As such, Eqn. \eqref{eqn:MI_entY} simplifies to $I(\bm{X};\bm{Y}) = h(\bm{Y})$. This means that for our mechanical encoder, a high entropy of $h(\bm{Y})$ results in a high mutual information $I(\bm{X};\bm{Y})$. In the case where $\bm{Y}$ captures all the information about the load, the conditional entropy $h(\bm{X}|\bm{Y}_s) = 0$ by definition. As a result, from Eqn. \eqref{eqn:MI_entX}, $\max I(\bm{X};\bm{Y}) = h(\bm{X})$. We use these observations to formulate the normalized mutual information for our mechanical encoder as:

\begin{equation}
\label{eqn:nmi}
NMI = \frac{I(\bm{X};\bm{Y})}{h(\bm{X})}
\end{equation}

Crucially, $NMI \in [0,1]$, which makes it easier to interpret when the mutual information between sensor readings and applied load $I(\bm{X};\bm{Y})$ is approaching the theoretical limit $h(\bm{X})$. Normalized mutual information will be used in Section \ref{result:architected} and \ref{result:optimize}. A simple numerical example to validate these bounds is demonstrated in Appendix \ref{appendix:info_bounds}.

\subsubsection{Construction of the Space of Applied Mechanical Loads}
\label{meth:load}

In this Section we outline our strategy for constructing the space of applied loads $\mathcal{X}$. Crucially, the load sample space $\mathcal{X}$ must give us insight into the role that mechanics plays in information transmission. From fundamental mechanics, Saint-Venant's principle states that the effects (i.e., displacement, strain and stress fields) from statically equivalent loads become indistinguishable at a sufficiently large enough distance from the applied load. In terms of information theory, we expect that if $\mathcal{X}$ is a space of statically equivalent loads, then $I(\bm{X};\bm{Y}) \rightarrow 0$ as the distance to the applied load increases. Accordingly, we construct two separate sets of $\mathcal{X}$ to encompass both statically equivalent loads as well as a larger class of loads to study additional mechanisms affecting mechanical information transmission.

To this end, we construct the space of applied traction $\mathcal{X}$ using a truncated generalized Fourier Series such that the $m^{th}$ sample with $d_x$ number of coefficients $c^m_n$ and basis functions $P_n$ can be represented as:

\begin{equation}
\label{eqn:applied_traction}
t^m_N(\zeta) = \sum_{n=0}^{d_x-1} c^m_n P_n(\zeta)  \, ,
\end{equation}

To sample from this space, we simply prescribe the number of coefficients $d_x$, which essentially determines the size of the sample space, and then individually sample each coefficient $c^m_n$ from a specified distribution. As such, $\bm{X} \in \mathbb{R}^{d_x}$ where the $m^{th}$ sample of $\bm{x} = [c_0^m, c_2^m,\dots,c_{d_x-1}^m]$. In this work, we sample $c^m_n\sim\mathcal{U}(-10,10)$ where $\mathcal{U}$ is the uniform distribution. We use Legendre polynomials as our basis functions, where each $n^{th}$ degree basis function $P_n(\zeta)$ can be generated in a compact form via the Rodrigues formula such that \citep{oden2017applied}:

\begin{equation}
P_n(\zeta) = \frac{1}{2^nn!} \frac{d^n}{d\zeta^n}(\zeta^2-1)^n \,.
\end{equation} 

Finally, for the applied load to have width $a$, we use change of variables $\zeta = s/a$ to transform $\zeta \in[-1,1]$ (i.e., the domain of Legendre polynomials) into the domain of applied traction $s \in [-a,a]$ where the tractions are applied in the $x_2$ direction (i.e. $\bm{t}^m_{d_x} = t_{d_x}^m \, \hat{e}_2$). Note that in this work, $x_1$, $x_2$, and $x_3$ are the coordinates in Cartesian coordinates, while $\bm{x}$ with no subscripts denotes a sample of input signal $\bm{X}$. 

In this work, we consider two sets of loads to perform our numerical experiments: $\mathcal{X}_{full}$, the set of loads with fixed $F$ and unconstrained resultant moment, and $\mathcal{X}_{even}$, the set on loads fixed magnitude $F$ and zero resultant moment. Exploiting the properties of Legendre polynomials, the total load magnitude $\int_{-a}^a t^m_{d_x}(s) \, ds = F$ is enforced by prescribing $c_0 = F/2$, which yields the first term of the Legendre polynomial to be $c_0P_0 = F/2a$. With $c_0$ now fixed, the remaining coefficients $c_1 \dots c_{d_x}$ are sampled to generate loads parameterized by $d_x$ coefficients. To construct $\mathcal{X}_{even}$, we only sample the even modes in the Legendre polynomials (i.e., $P_{n_{even}}$ where $n_{even} = 2n$ ; $n = 1,2,3,\dots$), ensuring symmetric loads with resultant moment of zero. In this work, without loss of generality, we set the load magnitude $F=1$ and the load width $a=100$. 

\subsubsection{Rate-Distortion Theory for a Mechanical Encoder}
\label{meth:rate_distortion}
While the primary focus of this work is on the mechanical encoder, for completeness, we will briefly analyze the full information channel in Section \ref{result:rate_distortion}. Since the full information channel reconstructs the load $\hat{\bm{X}}$ (see Fig \ref{fig:meth_info}a), its efficacy is evaluated by the reconstruction quality. Inevitably, the reconstruction quality depends on the amount of information transmitted through the channel, where, following intuition, more information transmission will result in a more accurate reconstruction. This fundamental trade-off between amount of transmitted data and quality of reconstruction is known as rate-distortion theory \citep{cover1999elements, gray1989source}. More formally, for a given encoder-decoder information channel, the amount of transmitted information, called the rate, is given by:

\begin{equation}
\label{eqn:distortion}
\text{Rate} = I(\bm{X};\hat{\bm{X}}) \, .
\end{equation}

In this work, the quality of reconstruction, referred to as the ``distortion,'' is characterized by using the mean-squared-error, which is expressed as:

\begin{equation}
\text{MSE} = \frac{1}{n\cdot d_x} \sum_{i=1}^n \lVert \bm{x}_i-\hat{\bm{x}_i}\rVert_2^2 \, .
\end{equation}

where $n$ denotes the number of samples. Note that contrary to typical mean-squared-error formulations, our MSE formulation is normalized by $d_x$ (i.e., the dimensions of $\bm{X}$), which follows the conventions for multivariate rate-distortion theory \citep{gray1989source}. 

The rate–distortion curve quantifies the theoretical optimum trade-off between transmission rate and reconstruction distortion and can be interpreted as a Pareto front, meaning that no point in the curve can be improved without sacrificing the other quantity. While the rate-distortion represents the true theoretical maximum, it oftentimes is difficult to compute in practice, particularly for continuous sources. As an tractable alternative, the \emph{lower bound} of the rate-distortion curve, called the Shannon lower bound, is computed instead. For a uniform source with $d_x$ dimensions with squared-error distortion, the Shannon lower bound is \citep{gray1989source, shannon1959coding}:

\begin{equation}
R_{SLB}(D) = \max \left\{0, \frac{1}{2} \log \frac{6}{\pi e D} \right\} \, ,
\end{equation}

where $R_{SLB}$ is the rate of the Shannon lower bound, $e$ is Euler's number, and $D$ is the distortion. More details on deriving the Shannon lower bound are presented in Appendix \ref{appendix:rate_distortion}. In Section \ref{result:rate_distortion}, we compare the performance of our mechanical information channel to this lower bound. We note that while our current work considers a noiseless channel, our framework can be extended to noisy channels through established results such as the noisy-channel coding theorem, channel capacity, and rate-distortion theory~\citep{cover1999elements,gray1989source,shannon1949communication}.

\section{Results and discussion}

As a step towards benchmarking mechanical intelligence, the primary goal of this work is to introduce a quantitative framework for mechanical information propagation. We begin by physically validating the mechanical encoder introduced in Section \ref{meth:info_channel} by using the elastic halfspace. In Section \ref{result:halfspace}, we connect information propagation to Saint-Venant’s principle and show in Section \ref{result:architected} that, through greedy sensor selection, the theoretical maximum mutual information between applied loads and sensor readings can be achieved. We then briefly consider the whole information channel in Section \ref{result:rate_distortion}, where we evaluate the performance of the mechanical information channel using rate-distortion theory as introduced in Section \ref{meth:rate_distortion}. Finally, we investigate the design of architected materials for manipulating information propagation. In Section \ref{result:mi_geom}, we demonstrated how changes in the domain geometry can either enhance or impede information propagation and related these effects to load paths. We conclude in Section \ref{result:optimize} by using Bayesian optimization to both maximize and minimize information propagation through geometric design. The examples considered in this work are simple ``toy problems,'' and they are intended as a starting point for applying an information-theoretic framework to analyzing, benchmarking, and ultimately designing mechanically intelligent materials and structures.

\subsection{Information Transmission in an Elastic Halfspace}
\label{result:halfspace}

\begin{figure}[pt]
    \centering
    \includegraphics[width= \textwidth]{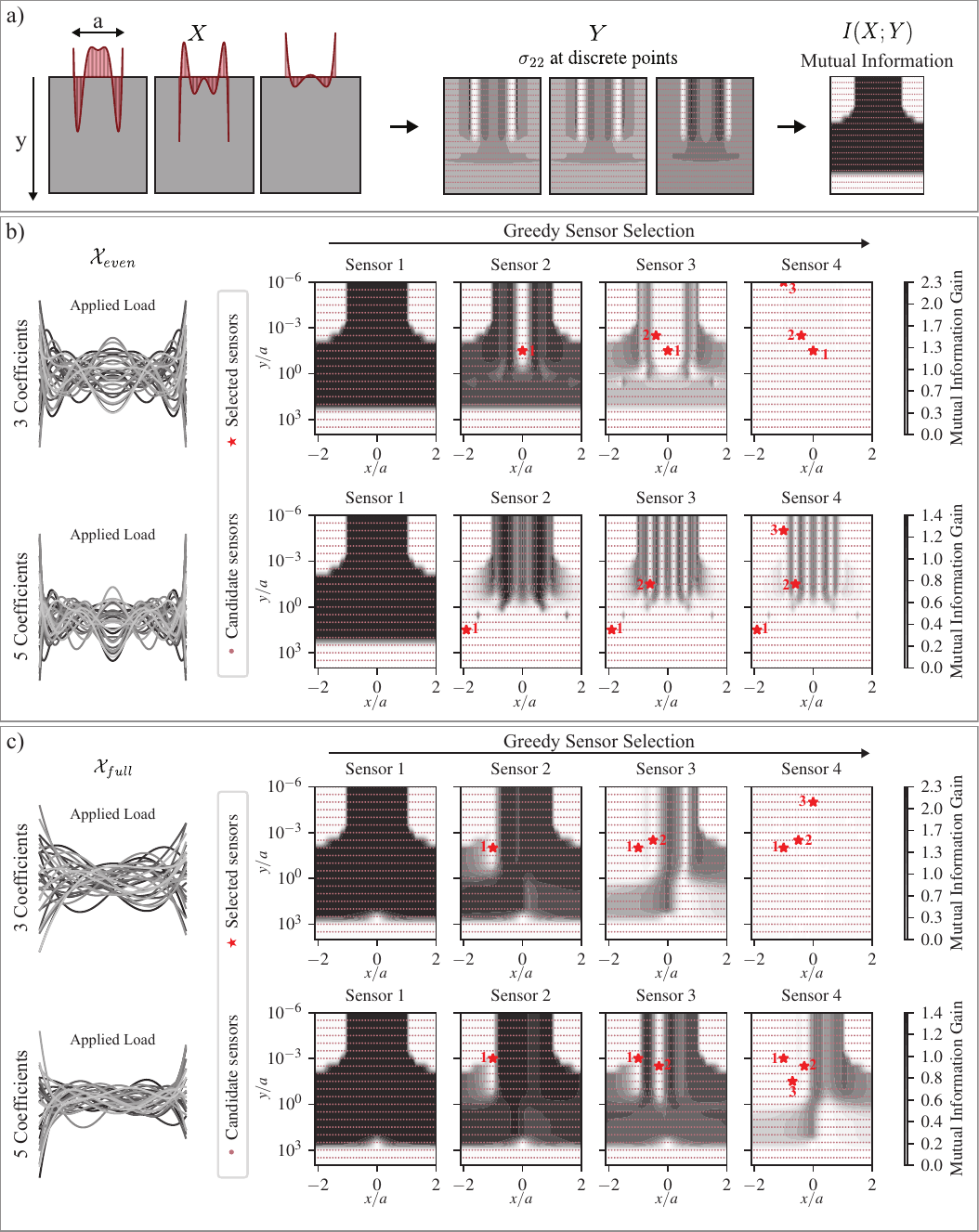}
    \caption{Elastic halfspace as an information encoder. a) Schematic of our pipeline to compute mutual information $I(\bm{X};\bm{Y})$ from input loads $\bm{X}$ and measured pointwise internal forces $\bm{Y}$. b) Visualizes pointwise changes of mutual information gain in the elastic halfspace subjected to $X\sim\mathcal{X}_{even}$ loads when sensors are sequentially greedily selected for $d_x = 3$ and $d_x = 6$. Note that $y/a$ is in log-scale but $x/a$ is in linear-scale. c) Visualizes changes of mutual information gain in the elastic halfspace subjected to $X\sim\mathcal{X}_{full}$ loads when sensors are sequentially greedily selected for $d_x = 3$ and $d_x = 6$. Note that $y/a$ is in log-scale but $x/a$ is in linear-scale.}
    \label{fig:half_space}
\end{figure}

We begin by validating our mechanical encoding framework on a simple testbed: an elastic halfspace used to examine how mechanical fields transmit information about applied loading. To mimic a practical physical system with sensors while preserving analytical simplicity, we measure reaction forces in the vertical direction $y$ (i.e. $\hat{e}_2$) at discrete points within the elastic body. We place sensors at $x \in[-2a,2a]$ at increments of $a/10$, and $y \in [10^{-4},10^6]$ at increments of $\sqrt{10}$ (i.e., half an order of magnitude). For simplicity, we assume that each sensor is small with respect to the elastic halfspace such that $\sigma_{22}$ is effectively constant over the area. This allows the reaction force measured at each discrete point to be represented directly as $\sigma_{22}$. In information-theoretic terms, the input signal $\bm{X}$ is the Legendre polynomial coefficients, and the output of the mechanical encoder $\bm{Y}$ (i.e., elastic halfspace) is the sensor readings. 

With this physical setup, we quantify information transmission under the set of applied loads $\mathcal{X}_{full}$ and $\mathcal{X}_{even}$ described in Section \ref{meth:load} to address two key questions: 1) Does information in elastic solids propagate in a way that is consistent with Saint-Venant's principle? and 2) Can the full information content of the applied load $\bm{X}$ be captured by discrete sensor measurements $\bm{Y}$? To address these questions, we follow the steps outlined in Fig. \ref{fig:half_space}a and apply both the even set of loads $\mathcal{X}_{even}$ and the full set of loads $\mathcal{X}_{full}$, each with width $a$, to the elastic half space. For each load set, $5000$ load realizations are sampled to estimate the mutual information. This sample size was found to be sufficient to obtain mutual information estimates. For each set of loads, the full stress field can be computed using linear superposition of the Flamant solution (see Appendix \ref{appendix:halfspace} for a detailed derivation). 

We first consider placing a single sensor in the elastic halfspace and computing the corresponding mutual information $I(\bm{X};\bm{Y})$ between the Legendre coefficients $\bm{X}$ and sensor measurements $\bm{Y}$. We repeat this computation for each candidate sensor locations in the elastic halfspace. The results of this analysis are visualized in column ``Sensor $1$'' in Fig. \ref{fig:half_space}b for $\mathcal{X}_{even}$ for $d_x=3$ (i.e., $3$ Legendre coefficients) and $d_x=5$ (i.e., $5$ Legendre coefficients), and in column ``Sensor $1$'' in Fig. \ref{fig:half_space}c for $\mathcal{X}_{full}$ for $d_x=3$ and $d_x=5$. First, we examine the relationship between Saint-Venant's principle and information propagation by comparing the results of $\mathcal{X}_{even}$ (Fig. \ref{fig:half_space}b) and $\mathcal{X}_{full}$ (Fig. \ref{fig:half_space}c). More specifically, in the $\mathcal{X}_{even}$ case (i.e., the set of statically equivalent loads), for both $d_x=3$ and $d_x=5$, we can observe that sensors at depth greater than $\sim 10^2$ provides negligible information about the applied load. This observation is consistent with Saint-Venant's principle. In contrast, for the $\mathcal{X}_{full}$ case, sensor readings remain informative about the applied load $\bm{X}$ up to depth of $\sim 10^3$, an increase of about $1$ order of magnitude in depth. In the $\mathcal{X}_{full}$ case, since the loads are not statically equivalent, the information decay is due to $\sigma_{22} \to 0$ as $y \to \infty$, a result from Flamant solution \citep{sadd2009elasticity}. 

\subsubsection{Mechanical Encoder with Greedy Sensor Selection Minimizes Information Loss}
\label{result:greedy_sensor}

Next, we investigate if discrete sensor readouts $\bm{Y}$ can approach the maximum mutual information about the applied load $\bm{X}$. 
To consider this question thoughtfully, we need a protocol for selecting sensor location. We aim to sequentially add discrete sensors from the set of candidate sensors $\bm{S}$ until subsequent sensor additions no longer provide information about applied load $X$. To evaluate the quality of the placed sensor, we use the conditional mutual information as a metric to evaluate mutual information gain, where for $k$ sensors the conditional mutual information is expressed as:

\begin{equation}
I(\bm{X};Y_{k}|Y_1 ,\dots, Y_{k-1}) = I(\bm{X};Y_1,\dots, Y_{k})-I(\bm{X};Y_1,\dots, Y_{k-1}) \, .
\end{equation}

As such, the conditional mutual information can be interpreted as the increase in mutual information from observing sensor $Y_k$ given the existing sensor set $\bm{Y}_s =\{Y_1, \dots, Y_{k-1} \}$. Note that in the event that $k=1$ (i.e., we are placing the first sensor), the conditional mutual information is equivalent to mutual information $I(\bm{X};Y_1)$. As shown in Algorithm \ref{algo:greedy_sensor}, we use the conditional mutual information to perform greedy sensor selection, where we sequentially select the candidate sensor $Y_i \in \bm{S}$ with the highest mutual information gain $I(\bm{X};Y_i|\bm{Y}_s)$. The selected sensor is then added to the existing set of sensors. The greedy sensor selection is terminated when the prescribed number of sensors $k$ is met.

\begin{algorithm}[H]
\label{algo:greedy_sensor}
\KwIn{Set of candidate measurements $\bm{S}$, number of sensors $k$, applied load $X$}
\KwOut{Set of selected measurements $\bm{Y}_s$}

Initialize $\bm{Y}_s \gets \{ \} $\;

\For{$i = 1$ \KwTo $k$}{
    Select
    $Y^\star = \displaystyle \arg\max_{Y_i \in \bm{S}}
    I(X; Y_i | \bm{Y}_s)$
    
    Add $Y^\star$ to $\bm{Y}_s$\;
}

\Return $\bm{Y}_s$\;
\caption{Greedy sensor selection via conditional mutual information}
\end{algorithm}

Fig. \ref{fig:half_space}b and Fig. \ref{fig:half_space}c plot the mutual information gain at candidate sensor locations $\bm{S}$ as we sequentially add sensors to the elastic halfspace via the greedy sensor selection strategy for applied loading from set $X\sim\mathcal{X}_{even}$ and $X\sim\mathcal{X}_{full}$ respectively. A visual comparison reveals that candidate sensors that are close to each other provide no mutual information gain, and the mutual information gain of candidate sensors is symmetric about $x=0$ for applied loads $X\sim\mathcal{X}_{even}$. In addition, for the case where $d_x=3$ (i.e. applied loads parameterized with $3$ Legendre coefficients), $3$ greedily selected sensors are sufficient to get maximum mutual information about the applied load since the $4^{th}$ candidate sensor provides no mutual information gain. Notably, this observation holds regardless of whether the set of loads is $\mathcal{X}_{even}$ or $\mathcal{X}_{full}$ (see the first row of Fig. \ref{fig:half_space}a and Fig. \ref{fig:half_space}b). However, for $d_x=5$, $4$ sensors are not enough to get the full mutual information about the applied load $X$ (see the second row of Fig. \ref{fig:half_space}a and Fig. \ref{fig:half_space}b). As such, in this study, we observe that $k = d_x$ (i.e., the number of sensors equals the number of Legendre coefficients used to parameterize the applied load $\bm{X}$) is sufficient for the greedily selected sensors to achieve  maximal mutual information encoded in the elastic halfspace about the applied load $\bm{X}$. We provide a more concrete numerical example of the greedy sensor selection approaching the theoretical maximum mutual information in Appendix \ref{appendix:info_bounds}.

To complement these numerical experiments, the results presented above can also be interpreted through standard linear algebraic arguments. We start by writing the mapping from the applied load $\bm{t}$ to the stress response $\bm{\sigma}(\bm{x})$ through a forward map $\bm{\mathcal{F}}$ where:

\begin{equation}
\label{eqn:sig_map}
    \bm{\sigma}(\bm{x}) = \bm{\mathcal{F}}(\bm{t}; \bm{x}),
\end{equation}
Discrete sensor readings $\bm{Y} \in \mathbb{R}^k$ at locations $\{\bm{x}_1,\dots,\bm{x}_k\}$ are obtained by the measurement operator $\bm{M}$, where:

\begin{equation}
\bm{Y} = \bm{M}\bm{\mathcal{F}}(\bm{t}), \qquad Y_i = \sigma_{22}(\bm{x}_i) \, .
\end{equation}
The full input-to-sensor map, from the load coefficients $\bm{c}$ to 
the sensor readings $\bm{Y}$, is then 
$\bm{c} \mapsto \bm{M}\bm{\mathcal{F}}\bigl(\bm{t}(\bm{c})\bigr)$, where $\bm{t}(\bm{c})$ is the parameterization of Eqn.~\eqref{eqn:applied_traction}. Note that $\bm{M}$ is linear in any setting as it simply samples a field, while $\bm{\mathcal{F}}$ is only linear when the relation Eqn. \eqref{eqn:sig_map} is linear.

For our linear elastic halfspace system, since $\bm{\mathcal{F}}$ is linear with respect to traction $\bm{t}$, from Eqn. \eqref{eqn:applied_traction} the linear superposition of the stress response from the individual Legendre polynomials can be expressed as:

\begin{equation}
    \bm{\sigma}(\bm{x}) = \sum_{n=0}^{d_x-1} c_n \bm{\mathcal{F}}(P_n; \bm{x}) = \sum_{n=0}^{d_x-1} c_n\, \bm{\sigma}^{P_n}(\bm{x}),
\end{equation}
where $\bm{\sigma}^{P_n}$ is the resultant stress field from $n^{th}$ Legendre polynomial when $c_n =1$. The sensor readings $\bm{Y}$ are then related to the coefficient vector $\bm{c} \in \mathbb{R}^{d_x}$ by selecting the measurement operator $\bm{M}$ such that:

\begin{equation}
\label{eqn:sensor_map}
    \bm{Y} = \bm{M}\bm{\mathcal{F}}\bigl(\bm{t}(\bm{c})\bigr) = \bm{A}\bm{c}, 
\qquad A_{in} = \sigma_{22}^{P_n}(\bm{x}_i).
\end{equation}
From
Eqn.~\eqref{eqn:sensor_map}, recovering $\bm{c}$ from $\bm{Y}$ reduces to a linear inverse problem: the coefficients are uniquely recovered if and only if
$\mathrm{rank}(\bm{A}) = d_x$, which in particular requires $k \geq d_x$ sensors. The threshold $k = d_x$ observed in our noiseless MI-based greedy
sensor selection (Fig.~\ref{fig:half_space}) is consistent with this rank condition; the greedy procedure identifies exactly $d_x$ sensor locations whose corresponding rows of $\bm{A}$ are linearly independent.

Crucially, the factorization of the sensor map in Eqn.~\eqref{eqn:sensor_map} into the forward operator $\bm{\mathcal{F}}$ and the measurement operator $\bm{M}$ separates two distinct reasons why $\bm{A}$ could lose rank, each attributable to each operator. The first lies in the mechanics: for the linear elastic halfspace, $\bm{\mathcal{F}}$ is injective so the basis responses $\bm{\sigma}^{P_n}(\bm{x})$ are linearly independent fields. The second lies in $\bm{M}$: even though these fields are independent, poorly chosen sensor locations $\{\bm{x}_i\}$  may produce redundant rows in $\bm{A}$. In other words, $\bm{\mathcal{F}}$ is fixed by the physics, while $\bm{M}$ is a design choice where for this problem, any rank deficiency in $\bm{A}$ stems from $\bm{M}$ alone and can be mitigated through better sensor placement. When $\mathrm{rank}(\bm{A}) = d_x$, reconstruction quality can be further quantified by $\sqrt{\det(\bm{A}^T\bm{A})}$. Geometrically, this measures how strongly and independently the coefficients $c_n$ are expressed in the readings $\bm{Y}$. The quantity $\sqrt{\det(\bm{A}^T\bm{A})}$ is known as the D-optimality criterion in optimal experimental design~\citep{bhattacharya2026optimal}. A larger value corresponds to smaller overall uncertainty in the recovered $\bm{c}$ under the presence of measurement noise, which we note is not considered in our numerical experiments. This use of mutual information for sensor placement also parallels information-theoretic approaches in structural health monitoring~\citep{papadimitriou2004optimal}, where the conditional entropy is minimized to reduce uncertainty.

\subsubsection{Sensor Readings with Maximum Mutual Information Can Reconstruct Applied Load}
\label{result:rate_distortion}
\begin{figure}[ht]
    \centering
    \includegraphics[width= 0.5\textwidth]{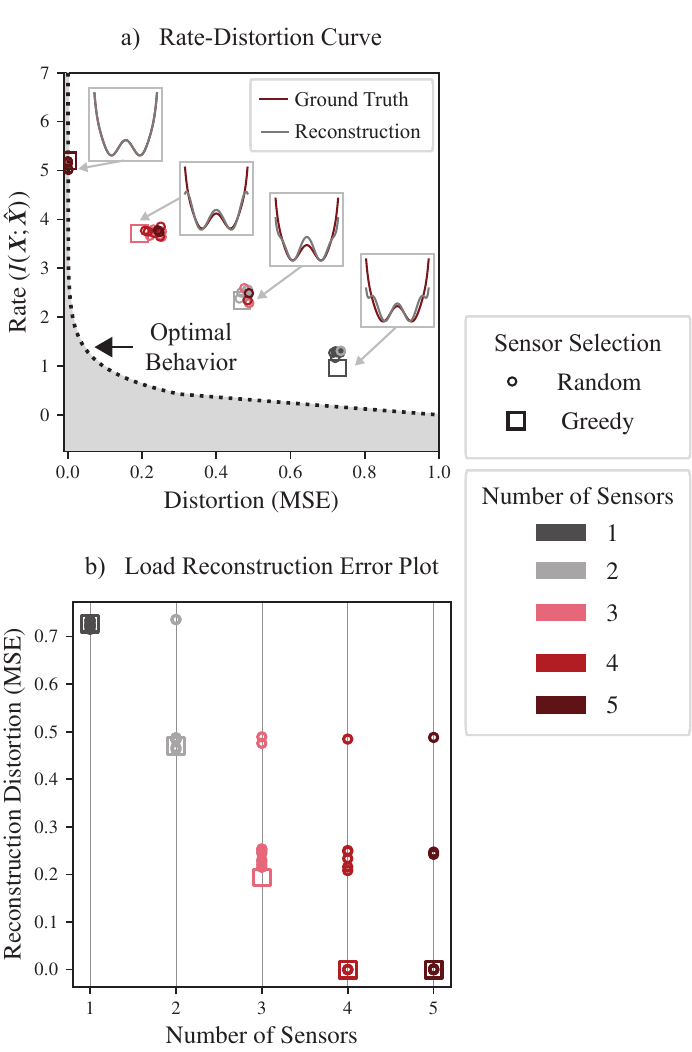}
    \caption{Elastic halfspace as an information encoder for a full information channel for $\mathcal{X}_{even}$ loads with $4$ Legendre coefficients ($d_x=4$). a) Rate-distortion curve of different number of sensors with different sensor selection schemes. The dotted line is represents the Shannon Lower Bound which is the theoretical lower bound of the optimal performance for a information channel. Insets show representative reconstructed load compared with ground truth for information channels with greedy sensor selection. b) Plot of raw data comparing the load reconstruction error (MSE) for different number of sensors with different sensor selection schemes.}
    \label{fig:rate_distortion}
\end{figure}

So far, we have shown that our greedy sensor selection scheme is able to approach the theoretical maximum $I(\bm{X};\bm{Y})/h(\bm{X})=1$. However, it remains unclear what achieving maximum mutual information implies from a practical perspective. Here we aim to link the notion of maximum mutual information with physical measurable quantities. To this end, we complete the information channel by introducing a decoder that performs load reconstruction $\hat{\bm{X}}$ from sensor readings $\bm{Y}$. Since the primary focus of this work is on constructing the mechanical encoder, to ensure minimal information loss between the encoded signal $\bm{Y}$ and reconstructed load $\hat{\bm{X}}$, we leverage the universal approximation theorem and use neural networks as our decoder. Details on the architecture of the neural network can be found at Appendix \ref{appendix:neural_network}.

Intuitively, a higher value of mutual information should lead to better load reconstruction because it implies that sensor measurements contains more information about the applied load. In Section \ref{result:greedy_sensor}, we demonstrated that with our greedy sensor selection scheme, increasing the number of sensors results in a higher mutual information. This naturally raises a fundamental question in rate-distortion trade-off question introduced in Section \ref{meth:rate_distortion}: given the amount of data transmitted from applied load $\bm{X}$ through the information channel, what is the resulting quality of signal reconstruction $\hat{\bm{X}}$? Note that the amount of data transmitted is controlled by the number of sensor readings $k$ (i.e., the dimensions of $\bm{Y}$).

We compare the performance of our mechanical information channel with the theoretical optimum derived in Section \ref{meth:rate_distortion} in Fig. \ref{fig:rate_distortion}a. The square markers denote information channel formed with greedy sensor selection, while the circular markers denote information channels formed from sensors randomly selected from $\bm{Y}_s$. The Shannon lower bound is represented by a dotted line, and the gray region is the theoretically unachievable region. As expected, the information channels follow the rate-distortion trade-off trend, with the greedy sensor selection scheme producing information channels that lie closer to the Shannon lower bound than those obtained with random sensor selection. In particular, in the cases $k \geq d_x$ (i.e., the number of sensors is greater than or equal to the number of Legendre coefficients), information channels from greedy sensor selection approach optimal performance, which is consistent with the notion that greedy sensor selection reaches maximum mutual information $I(\bm{X};\bm{Y})$. Complementing the notion that discrete sensor readings that approach maximum mutual information results in better load reconstruction, Fig. \ref{fig:rate_distortion}b shows that in general, increasing the number of sensors leads to a lower reconstruction error. Notably, information channels with greedy sensor selection are able to reach close to zero reconstruction error. We emphasize that zero reconstruction error is generally unattainable for continuous sources. We attribute this discrepancy between classical rate-distortion theory and our information channels to finite numerical precision. More comprehensive discussion on the discrepancy between rate-distortion theory and our mechanical information channel is found on Appendix \ref{appendix:rate_distortion}.

\subsection{Architected Materials Control Information Flow}

\begin{figure}[ht]
    \centering
    \includegraphics[width= \textwidth]{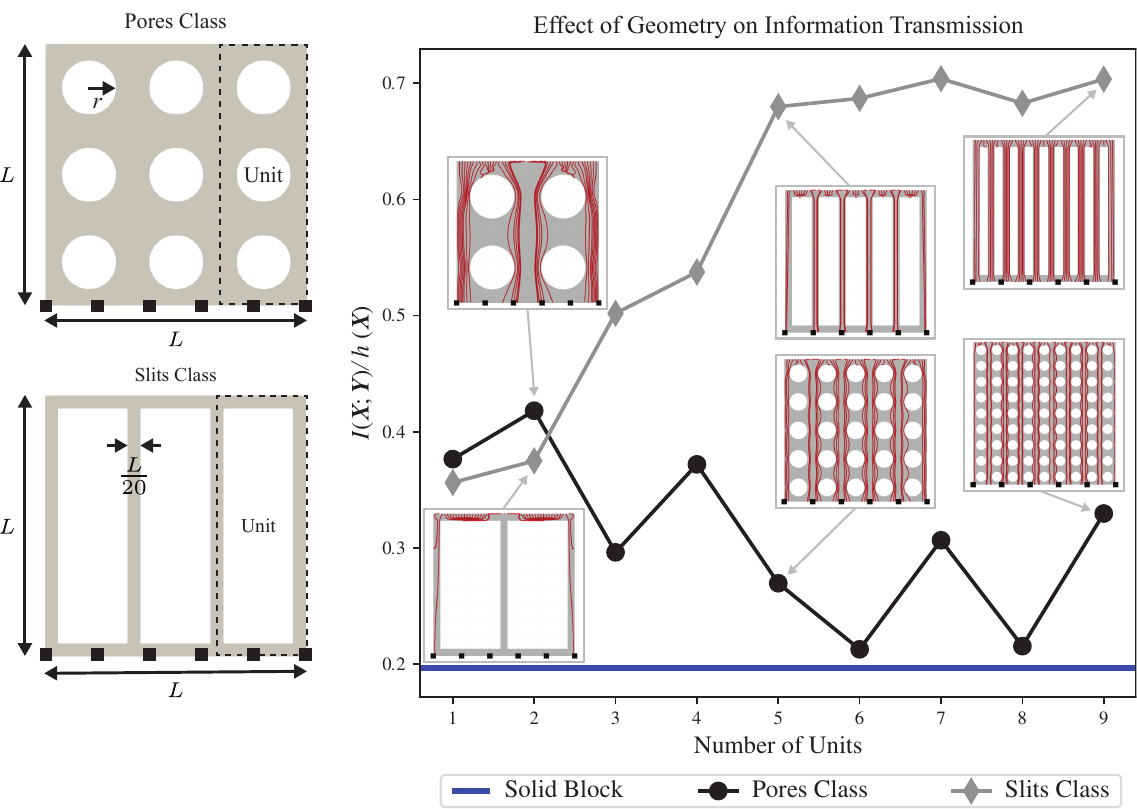}
    \caption{Effect of domain geometry on information transmission. The left panel visualizes a schematic of the geometries investigated here. Each dotted box contains one unit of the geometry class. Right panel plots the change of normalized mutual information $I(\bm{X};\bm{Y})/h(\bm{X})$ by changing the number of units for both the geometry classes. Insets show the geometry with principal stress lines traced to visualize information flow.}
    \label{fig:mi_geom}
\end{figure}

In Section \ref{result:halfspace}, we establish that a simple mechanical system, an elastic halfspace with sensors, can act as a mechanical information encoder. Here, we consider another toy problem where domain size and sensor locations are fixed, and examine how changes in domain geometry influence information transmission. We explore $2$ classes of material architecture: a ``pores class'' and a ``slits class.'' Both classes of geometry have width and height of $L$ (see left panel of Fig. \ref{fig:mi_geom}) and are fixed at the bottom. $\mathcal{X}_{full}$  with $d_x=6$ (i.e., $6$ Legendre coefficients) is applied to the top of the structure as load. Similar to the halfspace problem, we measure reaction forces in the $\hat{e}_2$ direction. We place $6$ evenly spaced sensors along the fixed boundary (i.e., $k=6$). We assume linear elastic behavior with $E=100$ and $\nu = 0.0$. We compute the reaction forces using FEniCSx, an open-source finite element software \citep{AlnaesEtal2014, baratta2023dolfinx, BasixJoss}. Details of our finite element implementation are presented in Appendix \ref{appendix:fea}. We parameterized the pore class by the porosity $\phi$ and number of units $n$. Each pore is circular with unit width $L_0=L/n$. As such, for a given porosity $\phi$, $r_0=(L_0\sqrt{\phi})/\sqrt{\pi}$~\citep{overvelde2012compaction}. For this study, we fix $\phi=0.3$ and only vary $n$. The slits class is parameterized only by number of units $n$, since the columns and connecting beams in the domain are fixed at $L/20$. Prior to running our investigation, we performed a mesh refinement study that is presented in Appendix \ref{appendix:mesh_refine}.

\label{result:mi_geom}
\label{result:architected}

We evaluate the efficacy of information transmission using the normalized mutual information $I(\bm{X};\bm{Y})/h(\bm{X})$ as shown in Eqn. \ref{eqn:nmi}. As  a quick reminder, $I(\bm{X};\bm{Y})/h(\bm{X}) = 0$ denotes no information transmission, while $I(\bm{X};\bm{Y})/h(\bm{X}) = 1$ corresponds to the maximum theoretical normalized mutual information for our mechanical encoder (see Appendix \ref{appendix:info_bounds} for details on the theoretical maximum mutual information). Note that, in this case, the theoretical maximum mutual information is achievable since the number of sensors is equal to the number of Legendre coefficients (i.e., $k = d_x$). Fig.\ref{fig:mi_geom} compares the normalized mutual information computed with $5000$ load realizations for both the pores class and the slits class of architected materials. Additionally, as a baseline comparison, we show the normalized mutual information of a solid domain size $L\times L$. For the slits class, increasing the number of units up to $5$ increases the normalized mutual information. In contrast, increasing the number of units for the pores class results in a  decreasing trend in the normalized mutual information. Notably, even though the pores class initially has higher normalized mutual information, after $2$ units, the pore class has a lower normalized mutual information that approaches the normalized mutual information of a solid block. 

The results from Fig. \ref{fig:mi_geom} demonstrate that modifications to the domain geometry can significantly affect information transmission. However, specific physical mechanisms linking domain geometry to information transmission are not reflected in the normalized mutual information. To gain insight, we examine load paths as a means of visualizing mechanical information flow. Specifically, we utilize principal stress lines \citep{kwok2016structural,yan2025principal} to visualize load paths. Details on how we generated principal stress lines are found in Appendix \ref{appendix:psl}. By visualizing principal stress lines for different architected geometries, we observe that structures in which principal stress lines connect the loaded boundary to the sensor locations exhibit higher normalized mutual information. This qualitative correlation suggests a link between load-paths and information propagation. While the precise quantitative relationship remains unresolved, these observations motivate future work aimed at establishing a predictive connection between load paths and mechanical information encoding.

\subsection{Maximization and Minimization of Information Transmission}
\label{result:optimize}

\begin{figure}[ht]
    \centering
    \includegraphics[width= \textwidth]{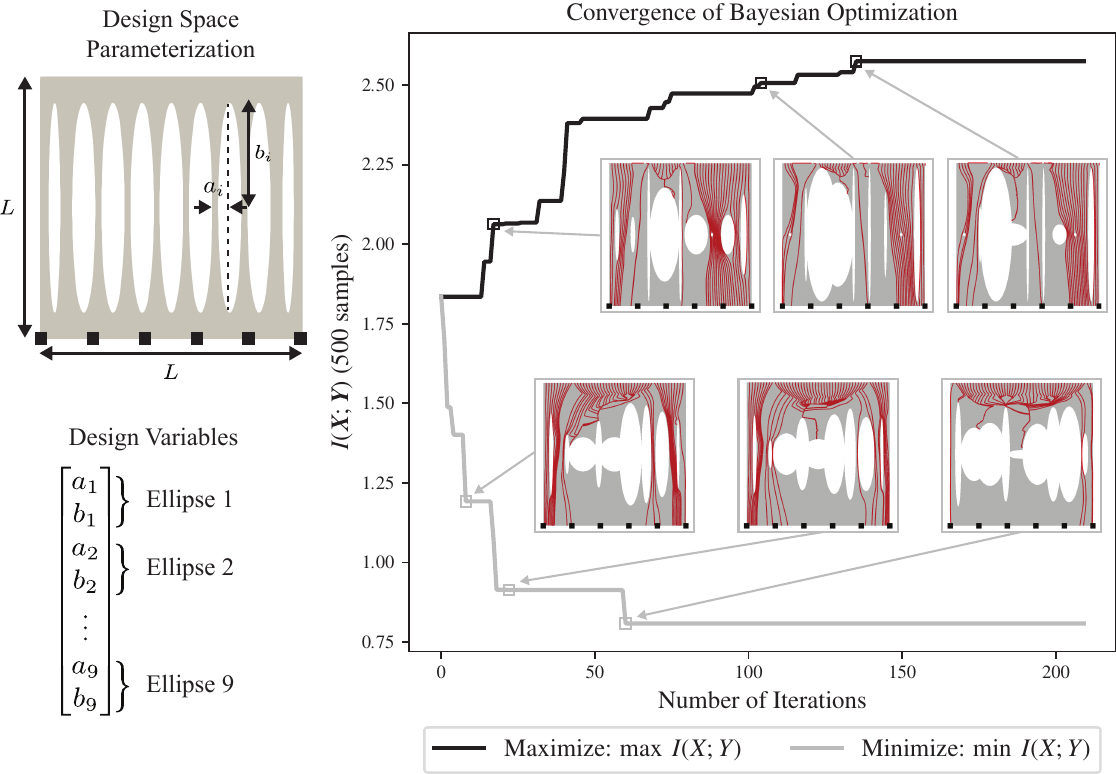}
    \caption{Bayesian optimization to maximize and minimize information transmission. a) Left panel visualizes design space parameterization for Bayesian optimization. Each individual $i^{th}$ ellipse is parameterized by $a_i$ and $b_i$. Right panel visualizes results from the Bayesian optimization. The black line shows the convergence plot during the maximization process, while the gray line shows the convergence plot during the minimization process. Insets visualizes the principal stress lines of the structure at different points in during the optimization process.}
    \label{fig:bayes_opt}
\end{figure}

Section \ref{result:mi_geom} showed that variations in the domain geometry of architected materials can influence information propagation in elastic solids. Here, we extend the idea even further and try to optimize a domain geometry for information transmission. The primary challenge in this optimization problem is the computation of mutual information $I(\bm{X};\bm{Y})$. Estimating mutual information requires approximating continuous probability distributions, which in turn demands a large number of samples $\bm{X}$ and $\bm{Y}$. Because $\bm{Y}$ is obtained from finite element simulations, generating each sample of $\bm{Y}$ can become computationally expensive. This cost becomes more pronounced in an optimization setting since many evaluations of of $I(\bm{X}; \bm{Y})$ are required. To mitigate this issue, we estimate $I(\bm{X};\bm{Y})$ with $500$ different samples at each optimization iteration instead of $5000$ samples like the previous halfspace examples, and treat the resulting mutual information estimate as a noisy objective. We also employ Bayesian Optimization as our optimization framework since it is well suited for objective functions that are expensive to evaluate and can tolerate noisy function evaluations \citep{brochu2010tutorial, frazier2018tutorial}. The constrained Bayesian Optimization framework is implemented with the Python package \verb|Bayesian Optimization| \citep{gardner2014bayesian, bayesopt}. Additional details on the Bayesian Optimization are presented in Appendix \ref{appendix:opt}.

The design space for the architected materials is parameterized by introducing $9$ elliptical voids into a square domain of size $L\times L$ (see Fig. \ref{fig:bayes_opt}). The shape of the $i^{th}$ ellipse is independently controlled by the major axis $b_i$ (vertical) and the minor axis $a_i$ (horizontal), where $i \in 1,\dots,9$ indexes the ellipses left to right. This parameterization results in $18$ independent design variables. To keep the domain connected from left to right, we constrain $b_i \in [L/100, 7L/15]$, ensuring a minimum $L/30$ distance between the ellipse and the top and bottom of the structure. To ensure connectivity from top to bottom, we let $a_1,a_9 \in [L/100, L/45]$ and $a_j \in [L/100, L/9]$ for $j = 2,\dots,8$. These constraints maintain a minimum distance $L/30$ at the domain boundaries, while allowing ellipses in the center to overlap. During the optimization process, for domain volume $V$, we constrain the void fraction $0.2 \leq V/L^2 \leq 1-9\pi/100$. Similar to Section \ref{result:mi_geom}, we place $6$ sensors (i.e., $k=6$) evenly spaced at the fixed bottom boundary and apply loads at the top from $\mathcal{X}_{full}$ parameterized by $6$ Legendre coefficients (i.e., $d_x = 6$). This configuration enables the mechanical encoder to reach the theoretical maximum mutual information $I(\bm{X};\bm{Y})/h(\bm{X})=1$.

Section \ref{result:mi_geom} demonstrates that modifying the domain geometry can either enhance or impede information transmission. Building on this observation, as a proof-of-concept, we aim to optimize the domain geometry to either maximize or minimize information transmission. Fig. \ref{fig:bayes_opt} presents the optimization convergence histories for both of these objectives (see Appendix \ref{appendix:opt} for raw optimization data). To get a more quantitative evaluation of the converged optimized structures, we compared the normalized mutual information (evaluated with $5000$ samples unseen during optimization) in Table \ref{tab:opt}. In the maximization case the optimized architected material marginally outperforms the best-performing slit configuration; however, the observed difference is small and might be attributed to variance in mutual information estimation. In the minimization case, the optimized design achieves values comparable to those of a solid block. However, none of the optimized structures clearly surpasses the extreme values identified in Section \ref{result:mi_geom}, nor do they approach the theoretical extrema. This gap suggests opportunities for future work, including the use of more expressive geometric parameterizations and more advanced optimization strategies, to design architected materials that more closely approach the theoretical limits of information transmission.

\begin{table}[h]
\centering
\begin{tabular}{cc}
\toprule
\textbf{Structure} &
\textbf{Normalized Mutual Information} \\
& (5000 samples) \\
\midrule
Bayesian Optimization  $\max I(\bm{X};\bm{Y})$ & 0.709 \\
Bayesian Optimization $ \min I(\bm{X};\bm{Y})$ & 0.202 \\
Slit geometry (maximum; $n=7$) & 0.704 \\
Pore geometry (minimum; $n=6$) & 0.213 \\
Solid block & 0.197 \\
\bottomrule
\end{tabular}
\caption{Normalized mutual information for different domain geometries.}
\label{tab:opt}
\end{table}

To visualize information propagation, we again show the principle stress lines for representative structures during the optimization process. Qualitatively, consistent with  Section \ref{result:mi_geom}, the architected material optimized for maximum information propagation has principle stress lines that propagate from the applied load on top of the domain to the sensor location. In contrast, for architected material optimized for minimum information propagation, the principle stress lines do not reach the sensors. In particular, the ellipses connect to form an effective barrier between the applied load and sensors, thereby deflecting the load paths away from the sensors. These observations further supports the notion that there is a strong connection between load paths and information propagation.

\changes{\subsection{A Preliminary Study of Information Transmission in a Nonlinear Mechanics Problem}
\label{result:elastica}}

\begin{figure}[pt]
    \centering
    \includegraphics[width=.9\textwidth]{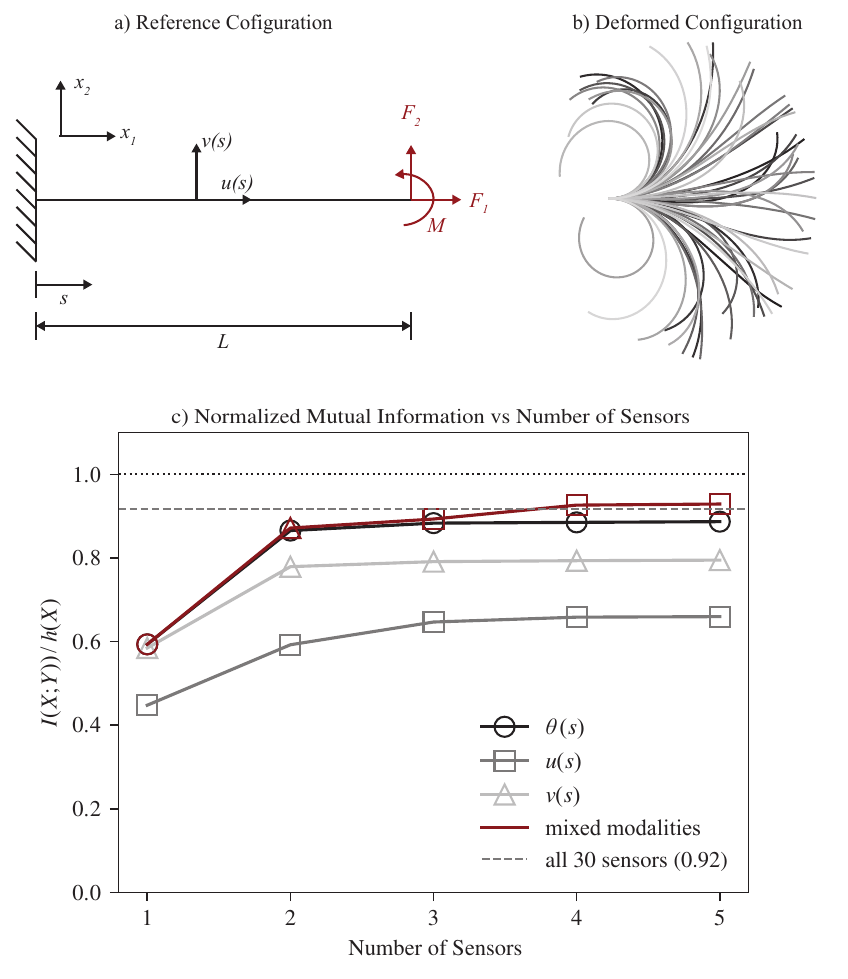}
    \changes{\caption{Visualization and results from the inextensible elastica problem introduced in Section \ref{result:elastica}. The elastica is parameterized by arclength $s\in[0,L]$ with $10$ evenly placed sensors along its body. a) Boundary conditions for our elastica problem. b) Representative solutions of the elastica problem with our loading parameters. c) Normalized mutual information $I(\bm{X},\bm{Y})/h(\bm{X})$ is compared to the number of sensors for different sensing modalities. Circular markers denote sensors measuring $\theta(s)$, square markers denotes sensors measuring $u(s)$, and triangular markers denote sensors measuring $v(s)$. Gray lines denote normalized mutual information from single modalities, while the red line denotes normalized mutual information from mixed modality (i.e., $\theta(s)$, $u(s)$, and $v(s)$). The vertical gray dotted line denotes the normalized information when all $30$ defined sensors are used.}
    \label{fig:elastica}}
\end{figure}

\changes{In Section \ref{result:halfspace}, we demonstrated that our mechanical encoding framework is applicable to simple, linear elastic systems. In this Section, we demonstrate preliminary results to confirm the feasibility of translating our mechanical encoding framework to nonlinear mechanical systems. As a simple representative problem, we consider an \textit{inextensible} elastica of length $L$ parameterized with arclength parameter $s\in [0,L]$~\citep{bigoni2015new, timoshenko2012theory}. The elastica is loaded with axial force $F_1$, shear force $F_2$, and an applied moment $M$ at $s=L$ while being clamped at $s=0$ (see Fig \ref{fig:elastica}a). As such, the $m^{th}$ input signal is $\bm{x} = [F^m_1, F^m_2, M^m]$. Each loading parameter is independently sampled from a uniform distribution where $F^m_1 \sim \mathcal{U}(-2,5)$, $F^m_2 \sim \mathcal{U}(-5,5)$, and $M^m \sim \mathcal{U}(-5,5)$. For simplicity, note that the range of $F_1$ is chosen to avoid buckling bifurcations by making sure the magnitude of the compressive load is below the critical buckling load. In this case, we allow the sensor measurements $\bm{Y}$ to have $3$ different modalities: the local orientation angle with respect to the $x_1$-axis $\theta(s)$, the horizontal displacement $u(s)$, and the vertical displacement $v(s)$. The measurements $\bm{Y}$ are taken at $10$ evenly placed sensors along the elastica. To be consistent with the halfspace examples from Section \ref{result:halfspace} in the manuscript, we use $5000$ samples to compute mutual information. Representative solutions from this loading parameterization and sampling can be found in Fig. \ref{fig:elastica}b. We defer interested readers to Appendix \ref{appendix:elastica} for detailed derivation and our numerical solution scheme to solve the elastica equation.} 

\changes{We perform the same greedy sensor selection scheme as demonstrated in Section \ref{result:greedy_sensor} and investigate if the conclusions formed from the elastic halfspace example carry over to the elastica problem. In particular, since the input signal $\bm{X}$ is parameterized by $3$ variables (i.e., $d_x = 3$), $k=3$ sensors should be sufficient to recover the full information content about the applied load according to the results in Section \ref{result:greedy_sensor}. In addition, we examine if the different measurement modalities affect the information content between $\bm{X}$ and $\bm{Y}$. We use the normalized mutual information $I(\bm{X};\bm{Y})/h(\bm{X})$ as shown in Eqn. \ref{eqn:nmi} to evaluate the efficacy of the elastica as an information encoder. Recall that the normalized mutual information requires $I(\bm{X};\bm{Y})/h(\bm{X}) \to 1$ as the sensors recover the full information content of the applied load (see Appendix \ref{appendix:info_bounds} for details). The elastica results are shown in Fig.~\ref{fig:elastica}c. For all cases, the normalized mutual information rises sharply from one to two sensors and then saturates, but for every measurement type it plateaus below $I(\bm{X};\bm{Y})/h(\bm{X}) = 1$. The $u(s)$ sensors are the least informative, while the $\theta(s)$, $v(s)$, and mixed cases perform comparably. Even using all available sensor measurements is insufficient to recover the full information content of the applied load.}

\changes{Currently, we do not have a definitive explanation for why the normalized mutual information $I(\bm{X};\bm{Y})/h(\bm{X})$ saturates below $1$ for this implementation of the elastica problem, but we believe it stems from the inextensibility constraint, which prevents $F_1$ from stretching the elastica. Instead, $F_1$ enters the equilibrium only through the moment it generates about deflected material points, with a lever arm set by the transverse deflection $v(s=L)$ of the elastica. When transverse deflections are small, this lever arm is small and $F_1$ barely contributes to the final deformed shape. This inextensibility constraint may contribute to $u(s)$ being the least informative sensor modality in Fig \ref{fig:elastica}c. In contrast, the transverse force $F_2$ and moment $M$ directly contribute to the deformation. As a result, loads that differ primarily in $F_1$ can produce nearly indistinguishable shapes, resulting in information about the axial load $F_1$ being weakly encoded in the elastica's final deformed shape. We note that this remains a hypothesis we have not tested directly, and we leave such an investigation to future work.}

\section{Conclusion}
\label{sec:conclusion}
In this work, we introduced a quantitative information-theoretic framework for analyzing information propagation and encoding in solid bodies. In our first result, we physically validate our premise of a ``mechanical encoder'' by using an elastic halfspace. Specifically, we link observed information propagation to Saint-Venant's effect, and demonstrated that a greedy sensor selection scheme can reach the theoretical optimal mutual information in our example system. We also complete the information channel by adding in a neural network based decoder and evaluated the performance of the whole information channel with rate-distortion theory. In our next set of results, we examine the potential for architected materials to influence mechanical information propagation by first investigating how changes in the domain geometry can affect information propagation, and how optimized architected materials can enhance and impede information propagation. \changes{We end by introducing preliminary work on extending our framework to nonlinear mechanics.} Overall, we find that our framework enables quantitative comparison and targeted design of systems with optimized mechanical information propagation.

We view this study as an initial step toward a quantitative, task-agnostic framework for mechanical intelligence, focused here on mechanical sensing. Important validation is still required to establish when the proposed measures reliably capture sensing performance across materials, geometries, and operating regimes. \changes{For instance, a natural next step is to thoroughly investigate the application of our information-theoretic approach to nonlinear mechanics. Our preliminary study demonstrates that this extension is non-trivial: even in a bifurcation-free regime, the sensor measurements were unable to recover the full information content of the applied load (achieving a maximum normalized mutual information of $0.92$ rather than $1.0$). Bifurcations, which we deliberately avoided in this initial study, may further break the one-to-one mapping between input and output and introduce additional complexity.} Moreover, information theory provides many candidates beyond Shannon entropy and mutual information, such as R\'enyi entropy and Kolmogorov complexity \citep{cover1999elements}, and identifying metrics that best reflect mechanically embodied information processing remains an open question. Finally, the framework should apply broadly across mechanically intelligent objectives. In this work, we define ``full information content'' in terms of accurate load reconstruction and rate-distortion theory. However, high reconstruction fidelity may not correspond to performance on other sensing objectives, nor to the information that matters for downstream decisions or control. Alternative formulations, such as the information bottleneck \citep{tishby2000information}, may therefore provide a more general basis for defining and optimizing mechanically relevant information. These alternative formulations are a promising direction for future investigation. 

We envision that the framework introduced in this work will provide a foundation for extending quantitative information-theoretic tools to other core components of mechanical intelligence beyond the sensing problems considered here. In brief, mechanical intelligence arises from the coordinated interaction of sensing, memory, and actuation, all of which can be viewed through the lens of information encoding, transformation, and utilization. Because this perspective requires only a joint distribution over an input and an output, and does not rely on linearity or static equilibrium, the framework established for sensing extends directly to memory and actuation, enabling systematic and quantifiable comparisons across systems and objectives. Mechanical memory has been realized through mechanisms such as multistability~\citep{jiao2023mechanical, yang2024mechanical}, path dependence~\citep{teunisse2025transition}, and non-reciprocal behavior~\citep{shaat2023chiral, yuan2025machine}. Within our framework, these correspond to taking the input to be a loading history and the output to be the state measured after the load has been removed. For example, in a multistable structure, the output is the final stable state, so the mutual information between loading and configuration is a function of the number of distinguishable stable states. In a path-dependent solid, the output is the residual strain field, while a purely elastic body returns to its reference configuration so that the retained information vanishes, consistent with it being memoryless. Similarly, while stimuli-responsive materials have demonstrated a wide range of mechanically embodied functions~\citep{xia2022responsive}, these behaviors are typically evaluated in task-specific terms. 
Specifically, for actuation, the input is a control signal and the output is the achieved configuration, such as the tip position of an actuator. Using our proposed framework, mutual information then measures the relationship between the actuation and the resulting configuration. Because mutual information depends only on the joint distribution of inputs and outputs and not necessarily on a one-to-one mapping between them, it still applies when a single input can produce several configurations, or several inputs the same configuration.
Future work will focus on leveraging information theory as a unifying quantitative language for integrating sensing, memory, and actuation, thus advancing a principled foundation for mechanical intelligence.

Concurrently, designing these systems will require robust computational tools, and we anticipate that a combination of topology optimization \citep{jia2024fenitop, sigmund200199} and data-driven methods \citep{hamdi2026towards, mohammadzadeh2023investigating, nguyen2024segmenting} will be essential for the exploration and optimization of mechanically intelligent systems. Finally, beyond engineered materials, this framework may also be adapted to study how mechanical information flows in biological systems during processes such as wound healing \citep{das2021extracellular}, growth and remodeling \citep{ambrosi2011perspectives, depalma2024matrix}, and cell–cell communication \citep{humphries2017mechanical, mann2019force}. As such, to ensure reproducibility and enable future extensions of this framework, we release the code used to reproduce all results in this work on GitHub under an open-source license. 

\section{Declaration of competing interest}
The authors declare that they have no known competing financial interests or personal relationships that could have appeared to influence the work reported in this paper.

\section{Acknowledgments}
This work was made possible with funding through the Boston University David R. Dalton Career Development Professorship, the Hariri Institute Junior Faculty Fellowship, the Haythornthwaite Foundation Research Initiation Grant, and the National Science Foundation Grants CMMI-2127864 and CMMI-2311640. This support is gratefully acknowledged. We also acknowledge the support of Boston University’s Research Computing Services for providing computing resources. 

\section{Additional Information}

All code, including the information-theory and finite-element analysis (FEA) components, is publicly available on GitHub (\url{https://github.com/pprachas/info_mech}
). Our information-theoretic estimation code is tested using \verb|pytest|, and common failure cases of the KSG estimator are also covered by the test suite.

\appendix

\section{A Primer on Information Theory}
\label{appendix:info_theory}
Information theory provides a mathematical framework for quantifying, transmitting, and storing information. For completeness within this manuscript, we introduce the key concepts from information theory that are necessary to fully understand this work. Note that we do not intend for this section to be a comprehensive overview of information theory. The concepts and explanations introduced in this Section are explained in more detail in Shannon's seminal work \citep{shannon1948amathematical, shannon1998themathematical}, as well as multiple educational resources such as ~\citep{cover1999elements, shannon1948amathematical}. Note that the information presented in this section is formulated for random variables; however, generalization to random vectors, as used in the main body of this work, is straightforward.

\subsection{Information Theory for Discrete Random Variables}
\label{appendix:dis_info}

At its core, information theory quantifies uncertainty of random variables through the use of \emph{entropy}. For a given discrete random variable $X$ from non-empty set $\mathcal{X}$ with a probability mass function $p(x)$ with $x \in \mathcal{X}$, the entropy $H(X)$ is defined as~\citep{cover1999elements, shannon1948amathematical}:
\begin{equation}
\label{eqn:entropy}
    H(X) = -\mathbb{E}_X [\log p(X) ] = -\sum_{x \in \mathcal{X}}p(x) \log p(x) \, ,
\end{equation}
where $\mathbb{E}_X[ \,\cdot \,]$ denotes the expectation operator.
The base of the logarithm determines the units. For log base $2$, entropy is expressed in \emph{bits}, while entropy is expressed in \emph{nats} for base $e$ (i.e., the natural log). In this work, unless explicitly mentioned, all logarithms will be the base of $e$. By convention $0\log0 = 0$ since $\lim_{x\to 0^+} \; x \log x = 0$. Entropy ranges between $ 0 \leq H(X) \leq \log|\mathcal{X}|$, where $|\, \mathcal{X}|$ denotes the set cardinality of $\mathcal{X}$, where $H(X) = \log|\mathcal{X}|$ if and only if $X$ has a uniform distribution over $\mathcal{X}$. Typically entropy can be interpreted as a measure of uncertainty. A high value of entropy for a random variable means that the random variable is hard to predict and contains a high amount of information. On the other hand, a random variable with $0$ entropy is deterministic and contains little information.

Entropy can also be extended to multiple variables. We introduce another random variable $Y$ from non-empty set $\mathcal{Y}$. For the pair of random variables $(X,Y)$ with joint distribution $p(x,y)$ the joint entropy is defined as~\citep{cover1999elements, shannon1948amathematical}:
\begin{equation}
\label{eqn:joint_entropy}
H(X,Y) = - \mathbb{E}_{X,Y} [\log p(X,Y)] = -\sum_{x\in\mathcal{X}}\sum_{y \in \mathcal{Y}} p(x,y) \log p(x,y) \, .
\end{equation}

Similarly, the conditional entropy is given as:
\begin{equation}
\label{eqn:conditional_entropy}
H(X|Y) = - \mathbb{E}_{X,Y} [\log p(X|Y)] = -\sum_{x\in\mathcal{X}}\sum_{y \in \mathcal{Y}} p(x,y) \log p(x|y) = -\sum_{y \in \mathcal{Y}} p(y)H(X|Y=y) \, .
\end{equation}

From Eqn.~\eqref{eqn:entropy}, Eqn.~\eqref{eqn:joint_entropy}, and Eqn.~\eqref{eqn:conditional_entropy}, the chain rule for entropy can be formulated. In general, for $n$ random variables $X_1,X_2,\dots, X_n$ random variables drawn from probability density $p(x_1,x_2,\dots,x_n)$ the chain rule for entropy is expressed as:

\begin{equation}
\label{eqn:chain_rule}
H(X_1,X_2,\dots,X_n) = \sum_{i=1}^n H(X_i|X_{i-1},\dots, X_1) \, .
\end{equation}

For the case of two random variables $X,Y$, the chain rule simple becomes $H(X,Y) = H(X)+H(Y|X)$. 

While entropy is useful in quantifying the uncertainty of random variables, mutual information is often used to measure the amount of information shared between random variables. For random variables $X$ and $Y$ with joint distribution $p(x,y)$ and marginal distributions $p(x)$ and $p(y)$, the mutual information between the two random variable is defined as:

\begin{equation}
\label{eqn:MI}
\begin{aligned}
I(X;Y) &=  \mathbb{E}_{X,Y} \left[\log \frac{p(X,Y)}{p(X) \otimes p(Y)} \right] = D \left( p(X,Y) \,\big|\big| \, p(X) \otimes p(Y) \right ) \\
&= \sum_{x\in\mathcal{X}} \sum_{y \in \mathcal{Y}} p(x,y) \log \frac{p(x,y)}{p(x)p(y)} \, ,
\end{aligned}
\end{equation}

where $D(\, \cdot \,)$ denotes the relative entropy (i.e., Kullback-Leibler divergence), and $\otimes$ denotes the outer product. Crucially, $I(X;Y)$ can also be written in terms of entropy $H(X)$ and $H(Y)$, where:

\begin{subequations}
\label{eqn:MI_entropy}
\begin{align}
I(X;Y) &= H(X) + H(Y) - H(X,Y) \label{eqn:MI_entfull}\\
&= H(X)-H(X|Y) \label{eqn:MI_entropyX}\\
&= H(Y) - H(Y|X) \label{eqn:MI_entropyY} \, ,
\end{align}
\end{subequations}

where Eqn.~\eqref{eqn:MI_entropyX} and Eqn.~\eqref{eqn:MI_entropyY} are obtained by combining Eqn.~\eqref{eqn:chain_rule} with Eqn.~\eqref{eqn:MI_entfull}. Note that using Jensen's inequality on Eqn. \eqref{eqn:MI}, it can be shown that $I(X;Y) \geq 0$ with the $I(X;Y) = 0$ if and only if $p(x,y) = p(x)p(y)$ (i.e., when $X$ and $Y$ are independent and share no information between each other). As a result, a high value of $I(X;Y)$ means that there is high amount of information content shared between $X$ and $Y$. Additionally, from Eqn. \eqref{eqn:MI_entropy}, mutual information can be interpreted a \emph{information gain}, since it denotes the reduction in entropy (i.e., uncertainty) after observing another random variable. It can also be observed from Eqn. \eqref{eqn:MI_entropy} that $I(X;Y) = I(Y;X)$. 

\subsection{Information Theory for Continuous Random Variables}
\label{appendix:meth_diffentropy}
In this Section we extend the quantities for discrete random variables defined in Section \ref{appendix:dis_info} to continuous random variables. For a continuous random variable $X$ with a corresponding probability density function $f(x)$ such that $\int_{-\infty}^\infty f(x) = 1$, the differential entropy of $X$ is defined as:

\begin{equation}
h(X) = -\mathbb E_X[\log f(X)] = - \int_{\mathcal{X}} f(x) \log f(x) \, dx \, ,
\end{equation}
where $\mathcal{X}$ is the support set, which is defined as the set of all $x \in \mathbb{R}$ where $f(x) >0$.
Note that in this work, the differential entropy will be denoted with $h(X)$ while the entropy of a discrete random variable will be denoted with $H(X)$. Similarly, the joint differential entropy can be defined as:
\begin{equation}
h(X,Y) = -\mathbb{E}_{X,Y} [\log f(X,Y)] = -\int_\mathcal{Y} \int_\mathcal{X} f(x,y) \log f(x,y) \, dx \, dy \, .
\end{equation}

Finally, the differential conditional entropy can be defined as:

\begin{equation}
h(X|Y) = -\mathbb{E}_{X,Y}[\log f(X|Y)]=-\int_{\mathcal{Y}}\int_{\mathcal{X}} f(x,y) \log f(x|y) \,dx \,dy \, .
\end{equation}

with $f(x|y) = f(x,y)/f(y)$ as defined in probability theory~\citep{pishro2014introduction}.

While differential entropy seems like a natural extension of discrete entropy, differential entropy can be negative and is dependent on the coordinate system of the random variable \citep{cover1999elements}. This inconsistency can be observed by performing dimensional analysis on the differential entropy with random variable $X$. The normalization condition that enforces $\int_{\mathcal{X}} f(x) \, dx= 1$ also results in $f(x)$ having dimensions of $[x]^{-1}$, which cause dependencies on the coordinate system  of $x$ when computing $\log (f(x))$ \citep{shannon1948amathematical}. As a result, we use the limiting density of discrete points (LDDP) formulation to compute the differential entropy that is independent from the choice of coordinate system. Following previous work \citep{jaynes1963information, nagel2024accurate}, we introduce an ``invariant measure'' $m(x)$ and define the continuous formulation of entropy  as the limit of the discrete entropy (i.e., Eqn \eqref{eqn:entropy}) as the number of partitions $N_p \to \infty$. As a start, we define $m(x)$ as:

\begin{equation}
\label{eqn:m_x}
m(x) = \left[ \,\lim_{N_p \to \infty} \, N_p \text{Vol}(x) \,\right]^{-1} \, ,
\end{equation}

where $\text{Vol}(x)$ denotes the volume of the quantized space of $X$ such that the \emph{discrete} probability mass function $p(x)$ can be written in terms of probability density function $f(x)$ as:

\begin{equation}
\label{eqn:dis_cont}
p(x) = f(x)  \text{Vol}(x) \, .
\end{equation}

By combining Eqn. \eqref{eqn:m_x} and Eqn. \eqref{eqn:dis_cont}, the discrete probability mass function can be written in terms of the continuous probability density function as:

\begin{equation}
p(x) = \lim_{N_p\to \infty} \, \left[ \frac{f(x)}{N_pm(x)}\right] \, .
\end{equation}

As a result, the corrected continuous relative entropy is defined as:
\begin{equation}
h_c(X) = \lim_{N_p\to \infty} \, \left[ \,-\int_\mathcal{X} f(x) \log \frac{f(x)}{N_pm(x)} \, dx \,\right] \, .
\end{equation}

However, as $N_p \to \infty$ the above formulation contains an unbounded $\log N_p$ term. Intuitively, you would need infinite number of nats (or bits depending on log base) to describe random variable $X$ to infinite precision \citep{cover1999elements}. To deal with the unbounded $\log N$ term, the LDDP  continuous entropy $h_{LDDP}(X)$ is defined as \citep{jaynes1963information}:

\begin{equation}
\label{eqn:relative_ent}
h_{LDDP}(X) = \lim_{N_p \to \infty} \, \left[ h_c(X) - \log N_p \right] = -\int_\mathcal{X} f(x) \log \frac{f(x)}{m(x)} \, dx \, .
\end{equation}

Similarly, the differential joint entropy is coordinate dependent. To remove this dependence, the LDDP joint differential entropy can be defined such that:

\begin{equation}
h_{LDDP}(X,Y) = -\int_{\mathcal{Y}}\int_{\mathcal{X}} f(x,y) \log \frac{f(x,y)}{m(x,y)} \,dx \, dy \, ,
\end{equation}
where $m(x,y)$ is the invariant measure on the joint space of $X$ and $Y$.

Lastly, a coordinate-independent formulation for differential conditional entropy can be defined as:

\begin{equation}
h_{LDDP}(X|Y) = -\int_{\mathcal{Y}}\int_{\mathcal{X}} f(x,y) \log \frac{f(x|y)}{m(x|y)} \,dx\,dy \, ,
\end{equation}

where the conditional invariant measure is defined as $m(x|y) = m(x,y)/m(y)$. Finally, the mutual information between continuous random variables $X$ and $Y$ can be defined as:

\begin{equation}
\begin{aligned}
I(X;Y) &= \mathbb{E}_{X,Y}\left[ \log \frac{f(X,Y)}{f(X) \otimes f(Y)} \right]= D(f(X,Y) \Vert f(X) \otimes f(Y)) \\
&= \int_{\mathcal{Y}}\int_{\mathcal{X}} f(x,y) \log \frac{f(x,y)}
{f(x)f(y)} \,dx \, dy \, .
\end{aligned}
\end{equation}

Crucially, notice that mutual information is coordinate independent. However, to ensure that the relationships in Eqn. \eqref{eqn:MI_entropy} are held, the differential relative entropies and conditional relative differential entropies must be evaluated on a consistent coordinate system. To this end, we use a variant of the popular Kraskov-St\"{o}gbauer-Grassberger (KSG) estimator \citep{kraskov2004estimating, nagel2024accurate} to estimate the coordinate-independent entropies. More details on the KSG estimator and the variant used in this work can be found in the Appendix \ref{appendix:ksg}. In this work, for brevity, all differential entropy $h(X)$ is the LDDP coordinate-independent formulation unless specified.

\subsection{Estimating Entropy and Mutual Information}
\label{appendix:ksg}
Estimation of differential entropy and mutual information is a non-trivial task since the continuous joint and marginal densities have to be estimated with a finite number of samples. As a result, there are a myriad of estimators available in literature, all with distinct advantages and disadvantages~\citep{czyz2023beyond}. The simplest estimators construct an approximation of the joint and marginal densities through histogram binning. However, binning strategy affects the value of mutual information, and although many adaptive binning schemes exist \citep{cellucci2005statistical, daub2004estimating}, it is not clear which binning schemes produce the least biased estimate of mutual information. On the other hand, neural estimators have been demonstrated to perform well in multiple benchmarking cases \citep{belghazi2018mutual, czyz2023beyond}, but they require training a neural network which can be computationally expensive. In this work, as a balance between complexity and accuracy, we choose the popular Kraskov-St\"{o}gbauer-Grassberger (KSG) estimator as our mutual information estimator \citep{kraskov2004estimating}. In brief, the KSG estimator uses the joint space $Z=(X,Y)$ to estimate the entropies by computing the distance norms in the joint space such that:

\begin{equation}
\lVert z_i-z_j\rVert = \max\{\lVert x_i-x_j \rVert, \lVert y_i-y_j \rVert  \} \, ,
\end{equation}
where $\lVert\, \cdot \, \rVert$ denotes the $p$-norm. While any $p$-norm can be used, we follow the original implementation and use the $\ell_{\infty}$ norm to compute the distances in the joint space~\citep{kraskov2004estimating}. By denoting the $k^{th}$ nearest $\ell_{\infty}$ distance of the point $i$ as $\rho_i/2$, the number of points in the partitioned marginal spaces $n_{x,i}$ and $n_{y,i}$ is computed by:

\begin{equation}
\begin{aligned}
    n_{x,i} &= \left| \left\{ x_j: \lVert x_i-x_j \lVert < \frac{\rho_i}{2}, \, i \neq j \right\}\right|  \\ 
    n_{y,i} &= \left| \left\{ y_j:  \lVert y_i-y_j \rVert < \frac{\rho_i}{2}, \, i \neq j \right\}\right| \, .
\end{aligned}
\end{equation}

 The continuous entropies in the marginal space for $N$ data points with $k$ nearest neighbors are then estimated by:
 
\begin{equation}
\label{eqn:ksg_ent}
\begin{aligned}
     h(X) &= -\langle \psi(n_{x,i}+1) \rangle +\psi(N) + \log c_{d_x} + d_X \langle \log \rho_i \rangle \\
     h(Y) &= -\langle \psi(n_{y,i}+1) \rangle +\psi(N) + \log c_{d_y} + d_y \langle \log \rho_i \rangle \, , 
\end{aligned}
\end{equation}

where $\langle \,\cdot \,\rangle = N^{-1} \sum_{i=1}^N (\cdot)_i$ for $N$ samples (i.e., the sample mean), $\psi(\,\cdot\,)$ denotes the digamma function, $d_x$ and $d_y$ are the dimensions of the $X$ and $Y$ spaces respectively, and $c_{d_x}$ and $c_{d_y}$ denotes the unit (hyper-)volume of the $X$ and $Y$ spaces respectively. Intuitively, the first two terms denotes the expected log-likelihood of the $k^{th}$ nearest neighbors with distance $\rho$, while the second term denotes the average log volume occupied by the points in the $k^{th}$ nearest neighbor. In the case of the $\ell_\infty$-norm the volume reduces to a cube with $c_{d_x} = c_{d_y} = 1$. Similarly, the joint entropy can be estimated as:

\begin{equation}
\label{eqn:ksg_jointent}
    h(X,Y) = -\psi(k) + \psi(N) +(d_X+d_Y) \langle \log \rho_i \rangle
\end{equation}

Subsequently, the mutual information is computed by combining Eqn. \eqref{eqn:ksg_ent} and eqn. \eqref{eqn:ksg_jointent} with Eqn. \eqref{eqn:MI_entropyfull} such that:
\begin{equation}
\label{eqn:KSG}
I(X;Y) = \psi(k) + \psi(N) - \langle \psi(n_x,_i+1)+ \psi(n_y,_i+1) \rangle
\end{equation}
Crucially, to be able to perform the simplification to get Eqn. \eqref{eqn:KSG}, the statistics must be computed in the joint space $Z=(X,Y)$. As a result, coordinate dependencies and estimation biases on $h(X)$, $h(Y)$, and $h(X,Y)$ cancels out and are thus not an issue our approximation of $I(X;Y)$. In this work, we will also compare $I(X;Y)$ with $h(X)$ and $h(Y)$. As a result, to make sure that the comparisons are meaningful, we take the differential entropy of random variables $X$ and $Y$ with the respective invariant measures $m(x)$, $m(y)$, and $m(x,y)$ with minimal changes to the KSG formulation such that the marginal and joint entropies are computed as \citep{nagel2024accurate}:

\begin{equation}
\label{eqn:ksg}
\begin{aligned}
     h_r(X) &= -\langle \psi(n_{x,i}+1) \rangle +\psi(N) + \log c_{d_X} + d_x \langle \log \rho_i \rangle + \langle \log m(x) \rangle \\
     h(Y) &= -\langle \psi(n_{y,i}+1) \rangle +\psi(N) + \log c_{d_y} + d_y \langle \log \rho_i \rangle + \langle \log m(y) \rangle \\
     h(X,Y) &= -\psi(k) + \psi(N) +(d_X+d_Y) \langle \log \rho_i \rangle + \langle \log m(x,y) \rangle \, .
\end{aligned} 
\end{equation}

Recall that we also want the errors in the invariant measures to cancel out such that the $I(X;Y)$ is unaffected by the invariant measure estimation since the value of $I(X;Y)$ is coordinate independent. As such, for simplicity, we follow \citep{nagel2024accurate} and choose the joint measure such that:

\begin{equation}
\label{eqn:factorize}
m(x,y) = m(x)m(y)
\end{equation}

and estimate the joint invariant measure with $k$-nn such that:

\begin{equation}
m(x,y) = [N_p\text{Vol}(x,y)]^{-1} = \left[ \frac{N}{k}c_{d_x}c_{d_y} \langle\rho^{d_x+d_y} \rangle \right] ^{-1}\, ,
\end{equation}

where the for $N$ samples $N_p=N/k$ (i.e., the number of partitions in the joint space), and $c_{d_x}c_{d_y}\langle \rho^{d_x+d_y} \rangle$ approximates the mean volume of a single data point. As a result, the marginal invariant measures are defined as:

\begin{equation}
\label{eqn:approx_m}
\begin{aligned}
    m(x) &= c_{d_x}^{-1}\left( \frac{N}{k} \langle \rho^{d_x+d_y} \rangle \right )^{-\frac{d_x}{d_x+d_y}} \\
    m(y) &= c_{d_y}^{-1}\left( \frac{N}{k} \langle \rho^{d_x+d_y} \rangle \right )^{-\frac{d_y}{d_x+d_y}}
\end{aligned}
\end{equation}

Finally, the coordinate-independent entropies are estimated by combining Eqn. \eqref{eqn:ksg} with eqn. \eqref{eqn:approx_m} to finally get \citep{nagel2024accurate}:
\begin{equation}
\label{eqn:rel_ksg}
\begin{aligned}
     h_r(X) &= -\langle \psi(n_{x,i}+1) \rangle +\psi(N) + d_x \langle \log \tilde{\rho}_i \rangle \\
     h_r(Y) &= -\langle \psi(n_{y,i}+1) \rangle +\psi(N) + d_y \langle \log \tilde{\rho}_i \rangle \\
     h_r(X,Y) &= -\psi(k) + \psi(N) +(d_x+d_y) \langle \log \tilde{\rho}_i \rangle \, ,
\end{aligned} 
\end{equation}

where $\tilde{\rho}_i = \rho_i/\langle \rho_i^{d_x+d_y} \rangle^{\frac{1}{d_x+d_y}}$. Note that due to the definition of continuous relative entropy, to deal with the unbounded $N_p \rightarrow \infty$ term (see Eqn. \eqref{eqn:relative_ent}), $N/k$ is subtracted from the LDDP entropy estimations to get eqn. \eqref{eqn:rel_ksg}.  We want to emphasize that eqn. \eqref{eqn:factorize} is an extremely restrictive condition that is true if and only if $m(x|y) = m(x)$ (i.e., when there are no relationship between $X$ and $Y$). However, by using this relation and Eqn. \eqref{eqn:MI_entropy}, it can be shown that computing $I(X;Y)$ with entropy relations (i.e., Eqn \eqref{eqn:MI_entropy}) will yield the same result as using the LDDP differential entropy formulation since the individual entropies are now coordinate-independent (i.e., the original KSG formulation to estimate mutual information does not change). On the other hand, this restriction furthermore induces an \emph{underestimation} of entropy due to errors of volume estimation between highly correlated variables, an issue that is also inherent in a traditional KSG estimator \citep{gao2015efficient}. This problem is numerically confirmed with our \verb|Pytest| suite found in our Github repository (\url{https://github.com/pprachas/info_mech}).  

In this work, to balance between estimator bias (i.e., large $k$) and variance (i.e., small $k$), all entropies and mutual information values are computed using $k=5$ nearest neighbors. Since the KSG estimator uses $k$-nn statistics, it is useful to scale $X$ and $Y$ to unit variance before using the KSG estimator \citep{kraskov2004estimating}. Scaling $X$ and $Y$ does not change the values of the LDDP entropies and mutual information since they are invariant under homeomorphism and are coordinate independent, but will help make sure that larger values will not dominate the $\ell_\infty$ norm when computing $k$-nn statistics. Here we use scikit-learn's \citep{scikit-learn} \verb|sklearn.preprocessing.StandardScaler| to transform the original data $X$ and $Y$ to $\hat{X}$ and $\hat{Y}$ such that both $\hat{X}$ and $\hat{Y}$ have unit variance. After scaling our data, we threshold individual values of $\hat{X}$ and $\hat{Y}$ with magnitudes that is less than $10^{-9}$ to $0$ in order to minimize numerical artifacts form machine precision.

\section{Arbitrary Load in an Elastic Halfspace}
\label{appendix:halfspace}

In this Section, we review and derive the stress distribution of a linear elastic isotropic half space under arbitrary loads. The derivation here is provided for completeness and can be found in educational resources in linear elasticity \citep{bower2009applied,sadd2009elasticity}. We consider an elastic halfspace that occupies a region $\Omega \in \mathbb{R}^3$ and boundary $\partial\Omega$. The displacement field is denoted with $\bm{u(\bm{x})} \equiv \bm{u}(x_1,x_2,x_3) $ with $\bm{x} \in \Omega$ being the spatial Cartesian coordinates. The linearized strain tensor $\bm{\varepsilon}$ in Cartesian coordinates is defined as:

\begin{equation}
\label{eqn:strain}
\varepsilon_{ij} = \frac{1}{2} \left[u_{i,j}+ u_{j,i} \right] \, .
\end{equation}

Note that we are using index notation using subscripts $i,j = 1,2,3$, with \emph{comma} in the subscript denoting partial differentiation such that $\frac{\partial}{\partial x_j}a_i = a_{i,j}$. The strain field of a continuous and single valued displacement field in simply connected bodies (i.e., strains that can be integrated to obtain displacements up to rigid body motion) must fulfill Saint-Venant compatibility conditions such that:
\begin{equation}
\varepsilon_{ij,kl} + \varepsilon_{kl,ij} - \varepsilon_{ik,jl} - \varepsilon_{jl,ik} = 0 \, .
\end{equation}

Next, the constitutive model for an isotropic linear elastic solid is denoted as:
\begin{equation}
\label{eqn:hooke}
\sigma_{ij} =  \frac{E}{1+\nu} \left\{ \varepsilon_{ij} + \frac{\nu}{1-2\nu}\varepsilon_{kk}\delta_{ij} \right\} ,
\end{equation}

where $\bm{\sigma}$ is the stress tensor, $E$ is Young's Modulus, $\nu$ is the Poisson's ratio, and $\delta_{ij}$ is the Kronecker delta. Again, note that due to index notation rules, summation is implied for repeated indices of the same term (i.e., $\sum_k \varepsilon_{kk} = \varepsilon_{kk}$). This relationship can be inverted to give the strain relationship with respect to stress. The equilibrium conditions for a quasistatic solid are:

\begin{equation}
\label{eqn:equilibrium}
\sigma_{ij,j} + b_i=0 \, ,
\end{equation}
with $\bm{b}$ denoting the body force. The boundary can be split into two portions where $\partial\Omega = \partial\Omega_u + \partial\Omega_\sigma$ with $\partial\Omega_u$ denoting the displacement boundary conditions and $\partial\Omega_\sigma$ denoting the traction boundary conditions such that:

\begin{align}
\label{eqn:bcs}
&& u_i &= q_i  \quad& \text{on} \; \partial\Omega_u && \\ 
&&  \sigma_{ij} \hat{n}_j &= h_i  \quad &\text{on} \; \partial\Omega_\sigma && 
\end{align}

where $q_i$ and $h_i$ are the prescribed displacement and traction boundary conditions respectively, and $\hat{\bm{n}}$ is the unit outward normal. 

\subsection{2D Elasticity}
\label{sec:meth_2D}

Since 3D elasticity problems are difficult to solve in general, we simplify the problem through plane approximations. The two common plane approximations are the \emph{plane stress} and \emph{plane strain} approximations. In general, the plane stress approximation is used for a body that is thin in the $\hat{e}_3$ direction with no out-of-plane forces. As such, the plane stress approximation yields:
\begin{equation}
\sigma_{13} = \sigma_{23} = \sigma_{33} = 0 \, .
\end{equation}

and the constitutive relations from Eqn.~\eqref{eqn:hooke} now reads:

\begin{equation}
\begin{aligned}
\sigma_{\alpha\beta} &= \frac{E}{1+\nu} \left\{  \varepsilon_{\alpha\beta} + \frac{\nu}{1-\nu}{\varepsilon_{\gamma\gamma}\delta_{\alpha\beta}} \right\}   \\
\varepsilon_{33} &= -\frac{\nu}{E} \left\{  \sigma_{11} + \sigma_{22} \right\} \, .
\end{aligned}
\end{equation}

Here we introduce \emph{Greek} subscripts for plane problems such that $\alpha, \beta,\gamma = 1,2$. Note that the stress-strain in the $\hat{e}_3$ directions are either zero, or can be written in terms of stress/strains in the $\hat{e}_1$ and $\hat{e}_2$ direction.

On the other hand, the plane strain approximation is most suited for a body that is long in the $\hat{e}_3$ direction, and as a result will have conditions:
\begin{equation}
u_3 = 0 \quad \text{and} \quad \varepsilon_{13} = \varepsilon_{23} = \varepsilon_{33} = 0 \, ,  
\end{equation}

with the constitutive relationship being simplified from Eqn.~\eqref{eqn:hooke} to:

\begin{equation}
\begin{aligned}
\varepsilon_{\alpha\beta} &= \frac{1+\nu}{E} \left\{ \sigma_{\alpha\beta} - \nu\sigma_{\gamma\gamma}\delta_{\alpha\beta}  \right\}
\\ 
\sigma_{33} &= \frac{E\nu  (\varepsilon_{11} + \varepsilon_{22})}{(1-2\nu)(1+\nu)} \,
\\
\sigma_{13}&=\sigma_{23}=0
.
\end{aligned}
\end{equation}
Again, note that the stress-strain in the $\hat{e}_3$ directions are either zero, or can be written in terms of stress/strains in the $\hat{e}_1$ and $\hat{e}_2$ direction. As such, for all plane conditions, the equilibrium condition as shown in Eqn.~\eqref{eqn:equilibrium} can be simplified to:

\begin{equation}
\label{eqn:equilibrium2D}
\sigma_{\alpha\beta,\beta} + b_{\alpha} = 0 \, ,
\end{equation}
and the Saint-Venant compatibility conditions can be simplified to:

\begin{equation}
\label{eqn:compatiblity2D}
\varepsilon_{11,22} -2\varepsilon_{12,12}+\varepsilon_{22,11}=0 \,.
\end{equation}

\subsection{Airy Stress Functions}
We then focus on the case where the body force can be written in a form where:

\begin{equation}
\label{eqn:airy_body}
b_\alpha=-V_{,\alpha}
\end{equation}
where $V$ is a scalar function of position. Combining Eqn.~\eqref{eqn:airy_body} with the equilibrium equations in Eqn.~\eqref{eqn:equilibrium2D}, the following relationship will hold:

\begin{equation}
\label{eqn:stress_airy}
\begin{aligned}
\sigma_{11} &= \varphi_{,22} + V \\
\sigma_{22} &= \varphi_{,11} + V \\ 
\sigma_{12} &= -\varphi_{,12}
\end{aligned}
\end{equation}
where $\varphi$ is a scalar potential function called the \emph{Airy stress function}. Note that the Airy stress function $\varphi$ automatically satisfies the equilibrium conditions by construction, while the compatibility conditions remain unsatisfied. As such, $\varphi$ must also fulfill (in direct notation):
\begin{equation}
\quad \nabla^4 \varphi = \mathcal{C} \nabla^2V \, ,
\end{equation}

where $\nabla$ is the del operator, making $\nabla^4$ the biharmonic operator and $\nabla^2$ the Laplacian operator, and $\mathcal{C}$ is the material parameters defined as\citep{sadd2009elasticity}:
\begin{equation}
\mathcal{C} =  \begin{dcases}
-(1-\nu) &\quad \text{for plane stress} \\
-\frac{1-2\nu}{1-\nu} &\quad \text{for plane strain}
\end{dcases}
\end{equation}
In the case of zero body forces, or in the case where the potential function satisfies $\nabla^2V=0$ the Airy stress function must satisfy:

\begin{equation}
\label{eqn:biharmonic}
\quad \nabla^4 \varphi = 0 \,.
\end{equation}

Note that in this case there is no distinction between the plane stress and plane strain conditions, and the solution is independent of all material parameters. As a result, the stress solution for a simply connected body with only traction boundary conditions will be identical, but the displacement and strain fields will be different due to their dependence on material properties (see Section~\eqref{sec:meth_2D}). In linear elasticity, Airy stress functions typically work best for problems with prescribed traction only (i.e., no prescribed displacements). Consequently, the stress and strain solutions will be unique, but the displacements can only be determined up to rigid body motion.

\subsection{Elastic Halfspace}
The approach to solve for the reaction force for an arbitrary applied load on an elastic halfspace is to first solve for the reaction forces for an applied line load $P$ in the $\hat{e}_2$ direction (i.e., Flamant's solution), before using linear superposition to compute the behavior for an arbitary load. For Flamant's problem, it is more convenient to work in polar coordinates; as a result, the domain of an elastic halfspace spans $r \in [0,\infty)$ and $\theta \in [0, \pi]$.  Next, the traction boundary conditions for Flamant's problems are:
\begin{equation}
\label{eqn:polar_bcs}
\sigma_{\theta\theta} = 0 \quad \text{and} \quad \sigma_{r\theta} = 0 \quad \text{on} \quad \theta = 0,\pi
\end{equation}

In addition, the relationship between stress and the Airy stress function in Eqn.~\eqref{eqn:stress_airy} in polar coordinates is (assuming no body force):

\begin{equation}
\label{eqn:polar_stress}
\begin{aligned}
\sigma_{rr} &= \frac{1}{r^2}\frac{\partial^2\varphi}{\partial\theta^2} + \frac{1}{r}\frac{\partial \varphi}{\partial r} \\ 
\sigma_{\theta\theta} &= \frac{\partial^2 \varphi}{\partial r^2} \\
\sigma_{r\theta} &= -\frac{\partial }{\partial r}\left(  \frac{1}{r} \frac{\partial \varphi}{\partial \theta}\right) \, .
\end{aligned} 
\end{equation}

To construct the solution, we first postulate that the solution is separable. Since we know that the line load $P$ has units of force per length, from dimensional analysis, we write the solution $\varphi$ to be in the form:
\begin{equation}
\label{eqn:phi_guess}
\varphi = Prf(\theta) \, , 
\end{equation}
where $f(\theta)$ is the unknown function to solve for. Eqn.~\eqref{eqn:phi_guess} is then combined with the (polar) biharmonic equation (Eqn.~\eqref{eqn:biharmonic}) to get:

\begin{equation}
\label{eqn:biharmonic_ode}
\frac{d^4f}{d\theta^4} + 2 \frac{d^2f}{d\theta^2} + f = 0 \, .
\end{equation}
The solution for this ODE is:

\begin{equation}
\label{eqn:ode_soln}
f(\theta) = C_1 \cos\theta + C_2 \sin \theta + C_3\theta\cos\theta + C_4\theta\sin\theta \, ,
\end{equation}

where $C_1$, $C_2$, $C_3$, and $C_4$ are unknown constants to be determined. By combining the ODE solution (Eqn.~\eqref{eqn:ode_soln}) with the Airy stress functions (Eqn.~\eqref{eqn:polar_stress}), the stress are written as:

\begin{equation}
\begin{aligned}
\sigma_{rr} &= \frac{2P}{r}\left( C_1\cos\theta - C_3\sin\theta \right) \\
\sigma_{\theta\theta} &= 0 \\
\sigma_{r\theta} &= 0 \, .
\end{aligned}
\end{equation}

Alternatively, the same guess can also be obtained by examining the General Mitchell's solution and selecting stress terms that scales with $1/r$ \citep{sadd2009elasticity}. Regardless, note that the stress terms $\sigma_{\theta\theta}$ and $\sigma_{r\theta}$ are automatically satisfied due to our choice of guess $\varphi$ (see Eqn.~\eqref{eqn:phi_guess}). In addition, the constant terms $C_2$ and $C_4$ do not contribute to stress, and can be set to $0$. The remaining constant can be determined by force balance of the radial forces in the $\hat{e}_1$ and $\hat{e}_2$ direction at arbitrary radius $R$ where:

\begin{equation}
\begin{aligned}
&\int_{0}^{\pi} \sigma_{rr}\cos(\theta) R \; d\theta = -C_1P\pi =  0  \\
&P-\int_{0}^{\pi} \sigma_{rr}\sin(\theta) R \; d\theta= P-C_3P\pi =0 \,.
\end{aligned}
\end{equation}
As a result, the radial stress in the halfspace is written as:
\begin{equation}
\sigma_{rr} = -\frac{2P}{\pi r} \sin\theta \, ,
\end{equation}
or alternatively, the reaction stress can be also written with respect to a Cartesian basis where:

\begin{equation}
\begin{aligned}
\sigma_{11} &= -\frac{2Px_1^2x_2}{\pi(x_1^2+x_2^2)^2} \\
\sigma_{22} &= -\frac{2Px_2^3}{\pi(x_1^2+x_2^2)^2} \\
\sigma_{12} &= -\frac{2Px_1x_2^2}{\pi(x_1^2+x_2^2)^2} \, .
\end{aligned}
\end{equation}

From superposition, for an arbitrary traction $t(s)$, the Airy stress function for the applied arbitrary load $\varphi$ can be written in terms of the Airy stress function of Flamant's Solution $\tilde{\varphi}$ as:
\begin{equation}
\varphi(x_1,x_2) = \int_{-\infty}^{\infty} t(s) \tilde{\varphi}(x_1-s,x_2)\ ds \, .
\end{equation}
Subsequently, the reaction stresses are expressed as:
\begin{equation}
\begin{aligned}
\sigma_{11} &= -\frac{2}{\pi} \int_{-\infty}^\infty {\frac{(x_1-s)^2x_2 \, t(s)}{\left((x_1-s)^2+x_2^2\right)^2}} \; ds \\
\sigma_{22} &= -\frac{2}{\pi} \int_{-\infty}^\infty {\frac{x_2^3 \, t(s)}{\left((x_1-s)^2+x_2^2\right)^2}} \; ds \\ 
\sigma_{12} &= -\frac{2}{\pi} \int_{-\infty}^\infty {\frac{(x_1-s)x_2^2 \, t(s)}{\left((x_1-s)^2+x_2^2\right)^2}} \; ds \, .
\end{aligned}
\end{equation}

We symbolically perform this integration using SymPy \citep{meurer2017sympy} to obtain $\sigma_{22}$. Because the formulation is expressed in terms of Legendre polynomials, we rely on SymPy’s polynomial integration routines. The resulting expressions involve complex logarithmic terms, which can be numerically unstable when evaluated directly; therefore, we rewrite the complex logarithms in terms of \texttt{atan2} to improve numerical robustness. In addition, the integration produces large rational polynomials whose evaluation over wide numerical ranges can lead to truncation errors under standard machine precision. To mitigate these issues, we use \texttt{mpmath} \citep{mpmath}, an arbitrary-precision numerical library, to evaluate the integrated expressions.

\section{Information-theoretic Bounds on Our Mechanical Encoder}
\label{appendix:info_bounds}
\begin{figure}[ht]
    \centering
    \includegraphics[width= \textwidth]{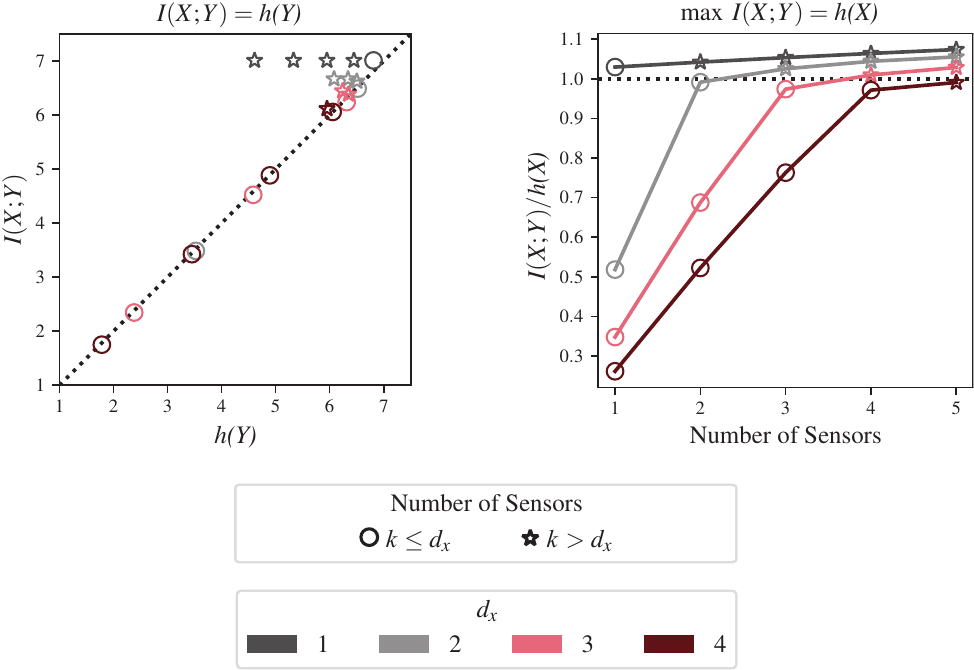}
    \caption{Numerical validation of information bounds using results from the elastic halfspace problem. Left panel demonstrates $I(\bm{X};\bm{Y})=h(\bm{Y})$ for most cases except for cases where the number of sensors $k$ is greater than the number of Legendre coefficients $d_x$. Note that the dotted line denotes the identity function. Right panel shows that $\max I (\bm{X};\bm{Y}) = h(\bm{X})$ for most cases with exceptions in cases where the number of sensors is greater than the number of Legendre coefficients $d_x$. Note that the dotted line here denotes the theoretical maximum $I(\bm{X};\bm{Y})/h(\bm{X}) = 1$.}
    \label{fig:info_bounds}
\end{figure}

This Section presents numerical studies to validate the information bounds derived at the end of Section \ref{meth:info_channel}. Additionally, since the numerical example are from in the same setting as elastic halfspace problem investigated in Section \ref{result:greedy_sensor}, this section also serves as numerical validation that the greedy sensor selection with $k=d_x$ reaches the maximum theoretical mutual information in our information channel.

We first start with the statement that for our noiseless mechanical encoder:
\begin{equation}
I(\bm{X};\bm{Y}) = h(\bm{Y}) \, .
\end{equation}

The left panel of Fig. \ref{fig:info_bounds} compares $I(\bm{X}, \bm{Y})$ with $h(\bm{Y})$ and shows that for $k \leq d_x$ (i.e., the number of Legendre coefficients is less than or equal to the number of sensors), $I(\bm{X};\bm{Y})=h(\bm{Y})$ since it coincides with the identity function. In the case where $k > d_x$, note that the value of $I(\bm{X};\bm{Y})$ remains the same as the case $k = d_x$, but $h(\bm{Y})$ decreases with increasing number of sensors. This is a strong indication that the discrepancy between $I(\bm{X};\bm{Y})$ and $h(\bm{Y})$ is due to the challenges associated with estimating $h(\bm{Y})$ and its invariant measure $m(\bm{Y})$ when there are strong correlations between $\bm{X}$ and $\bm{Y}$. Note that in our case the invariant measure is dependent on the joint space $(\bm{X},\bm{Y})$. More details on estimating the invariant measure and differential entropy is found in Appendix \ref{appendix:ksg}.

The second statement statement establishes bounds on the maximum mutual information for our mechanical encoder where:

\begin{equation}
\max \, I(\bm{X};\bm{Y}) = h(\bm{X}) \, .
\end{equation}

The right panel shows that that the $(\max I(\bm{X};\bm{Y})/h(\bm{X})$ approaches $1$ with increasing $d_x$ when $k < d_x$ and gets close to $1$, which is the ideal theoretical limit,  when $k = d_x$. However, when $k > d_x$, $(\max I(\bm{X};\bm{Y})/h(\bm{X})$ still increases, but at a slower rate than when $k < d_x$. Note that from the left plot, since $I(\bm{X};\bm{Y})$ does not change as in the case where $k=d_x$, the $I(\bm{X};\bm{Y})/h(\bm{X})$ comes from the decrease $h(\bm{X})$ due to high correlations in between $\bm{X}$ and $\bm{Y}$. The fact that $I(\bm{X}, \bm{Y})$ remains constant for 
$k>d_x$ provides further evidence that the mutual information has reached its maximum value.

\section{Additional Numerical Experiments for Greedy Sensor Selection}

The results in Section \ref{result:greedy_sensor} indicate that the condition $k=d_x$ is sufficient for the greedily selected sensors to achieve maximum mutual information. Additionally, the linear algebraic analysis at the end of the section suggests that this result holds even when the underlying mechanics problem changes. In this Section, we perform additional numerical experiments to further demonstrate and strengthen this claim.

\subsection{The Elastic Halfspace Problem with Legendre Coefficients from a Normal Distribution}
\begin{figure}[pt]
    \centering
    \includegraphics[width= \textwidth]{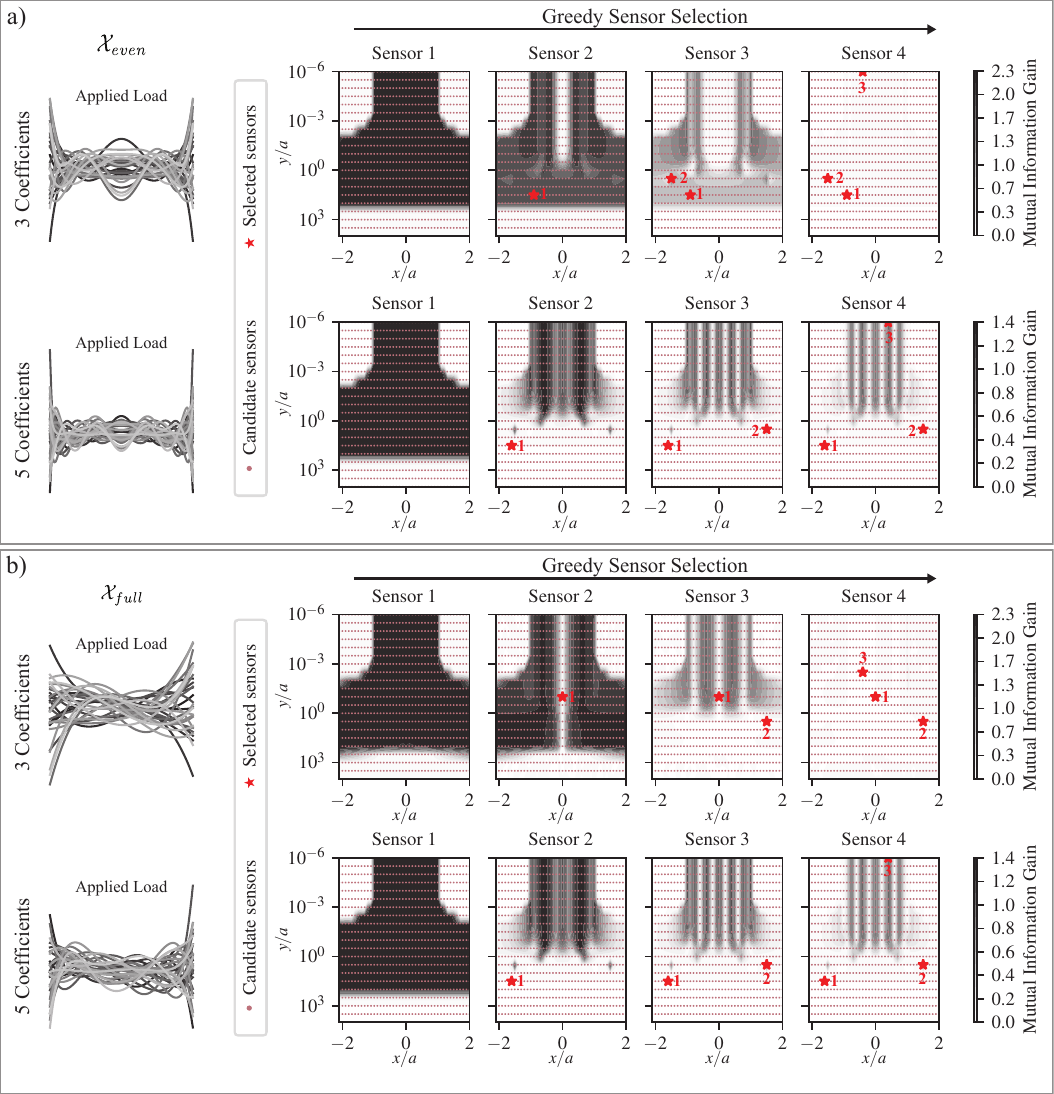}
    \caption{Elastic halfspace as an information encoder. The greedy sensor selection scheme is identical to Section \ref{result:greedy_sensor}, with the only difference being the sampling strategy of the Legendre coefficients, where $c_n^m \sim \mathcal{N}(0,1)$. a) Visualizes pointwise changes of mutual information gain in the elastic halfspace subjected to $X\sim\mathcal{X}_{even}$ loads when sensors are sequentially greedily selected for $d_x = 3$ and $d_x = 6$. Note that $y/a$ is in log-scale but $x/a$ is in linear-scale. b) Visualizes changes of mutual information gain in the elastic halfspace subjected to $X\sim\mathcal{X}_{full}$ loads when sensors are sequentially greedily selected for $d_x = 3$ and $d_x = 6$. Note that $y/a$ is in log-scale but $x/a$ is in linear-scale.}
    \label{fig:half_space_normal}
\end{figure}

We slightly modify the elastic half-space problem to examine whether the results in Section \ref{result:greedy_sensor} still hold. More specifically, the applied load is still parameterized by the Legendre coefficients defined in Eqn. \eqref{eqn:applied_traction}. However, instead of sampling the coefficients from a uniform distribution, $c_n^m \sim \mathcal{U}(-10,10)$, the coefficients are sampled from a standard normal distribution with zero mean and unit variance, i.e., $c_n^m \sim \mathcal{N}(0,1)$. The results are shown in Fig. \ref{fig:half_space_normal}a and Fig. \ref{fig:half_space_normal}b for cases $d_x=3$ and $d_x=5$ respectively. Notably, although the sensor locations differ from the previous case, the relationship between the number of sensors $k$ and the number of Legendre coefficients $d_x$ remains unchanged: when $k<d_x$, the sensors are unable to capture the full information content of the applied load, whereas $k=d_x$ is sufficient to achieve the maximum mutual information.

\subsection{The Elastic Halfspace Problem with Discontinuous Loading}

\begin{figure}[h]
    \centering
    \includegraphics[width= \textwidth]{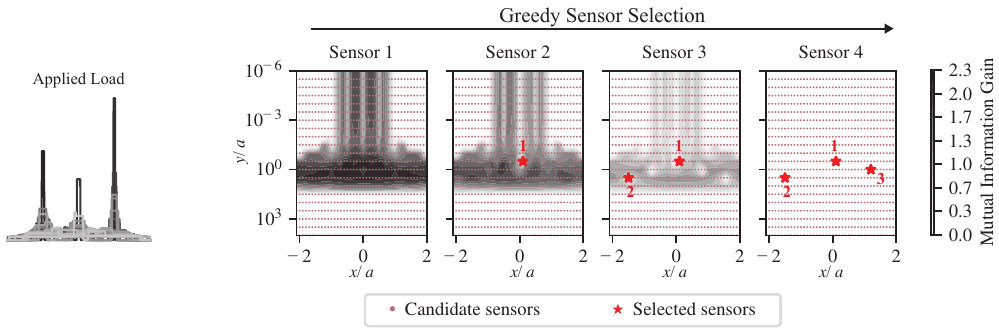}
    \caption{Elastic halfspace as an information encoder under discontinuous loading. The greedy sensor selection scheme is identical to Section~\ref{result:greedy_sensor}; the only difference is how the applied load is constructed. Here the traction over $[-a,a]$ is built from three uniform loads centered at $-a/2$, $0$, and $a/2$, each with width $w_i \sim \mathcal{U}(a/100, a/2)$ and magnitude $1/(6w_i)$ so that each contributes one third of the total force and the three sum to a unit total force $F=1$. Each column visualizes the pointwise mutual information gain in the elastic halfspace as sensors are sequentially added by greedy selection. Note that $y/a$ is on a log scale while $x/a$ is on a linear scale.}
    \label{fig:halfspace_discontinuous}
\end{figure}

In this Section, we slightly modify the elastic half-space problem to examine whether the results in Section \ref{result:greedy_sensor} continue to hold when the applied load is discontinuous over the subdomain $[-a,a]$. For this case, the total applied load is constructed from three uniform loads with centroids located at $-a/2$, $0$, and $a/2$. The width of each load, $w_i$ for $i=1,2,3$, is independently sampled from a uniform distribution, $w_i \sim \mathcal{U}(a/100,a/2)$. The magnitude of each load is fixed at $1/(6w_i)$ such that each load contributes $1/3$ of the total applied force, yielding a unit total force $F=1$. The solution is obtained through linear superposition of the responses to the three individual uniform loads. As shown in the top panel of Fig. \ref{fig:halfspace_discontinuous}, and consistent with the other half-space examples in Section \ref{result:greedy_sensor}, three sensors are sufficient to capture the full information content of the applied load.

\section{Additional Information on Rate-Distortion Theory}

\label{appendix:rate_distortion}
Rate-distortion theory is a branch of information theory that explores the optimal trade-off between rate (i.e., amount of information transmitted) and distortion (i.e., loss of quality) in an between the original signal $X$ and reconstructed signal $\hat{X}$ in an information channel. More formally, the distortion measure is a mapping $d(x,\hat{x}): \mathcal{X} \times \hat{\mathcal{X}} \to \mathbb{R}^+$. The distortion associated with the entire information channel is defined as:

\begin{equation}
D = \mathbb{E}[d(x,\hat{x})] = \int_\mathcal{X}\int_{\hat{\mathcal{X}}}p(x,\hat{x})d(x,\hat{x}) \, d\hat{x} \, dx = \int_\mathcal{X}\int_{\hat{\mathcal{X}}} p(x) p(\hat{x}|x)d(x,\hat{x}) \, d\hat{x} \, dx 
\end{equation}

where $D$ is the distortion, $p(x)$ is the probability density function of the original signal, $p(x,\hat{x})$ is the joint probability density function between the original signal and the reconstructed signal, and $p(\hat{x}|x)$ conditional probability density function of $\hat{X}$ having observed $X$. To formulate the rate-distortion of a set of input-output pair $(x,\hat{x}) \in (X,\hat{X})$, let $D^\star$ be the distortion associated with the information channel. The rate-distortion is defined to be:

\begin{equation}
R(D) \min_{D^\star \leq D} I(X;\hat{X}) \, .
\end{equation}

In other words, for a given distortion measure, the rate-distortion function minimizes that rate over all the possible information channels (i.e., the conditional distribution $p(x,\hat{x})$). As a result, note that the rate-distortion function is \emph{not} a property of a specific information channel, but a property of input $X$ and distortion measure. Note that the converse, where the rate is given and the minimum distortion is computed, is then called the distortion-rate function, and is equivalent to the rate-distortion function due to monotonicity and convexity of the rate-distortion function \citep{cover1999elements, gray1989source}.

Except for a few ideal cases, the rate-distortion curve is difficult to analytically solve. Typically, for discrete sources, the rate-distortion curve can be computed using the Blahut-Arimoto algorithm \citep{arimoto1972algorithm,blahut2003computation}, but estimating the rate-distortion curve for continuous sources is still an open research area. Alternatively, the lower bound of the rate-distortion curve, also known as the Shannon lower bound \citep{shannon1959coding}, is used to approximate the rate-distortion curve. 

We by deriving the rate-distortion function for our mechanical encoder by first starting with the case of $d_x=1$ (i.e., for loads represented by $1$ Legendre coefficient). In general, for information channels with input source $X\in\mathbb{R}$ a, the Shannon lower bound is expressed as:
\begin{equation}
R_{SLB}(D) = \max \{0,h(X) - h(g_s) \} \, ,
\end{equation}

where $g_s$ is related to the distortion measure, and the $\max$ operator ensuring $R>0$. More details on $g_s$ can be found in \citep{gray1989source}.
In our case with the mean-squared-error as the distortion function, the Shannon lower bound can be expressed as:

\begin{equation}
\label{eqn:SLB}
    R_{SLB}(D) = \max \{ 0,h(X) - \frac{1}{2}\log2eD \} \, .
\end{equation}

Since we scale our data with sklearn's \verb|sklearn.preprocessing.StandardScaler| \citep{scikit-learn}, the input loads $X$ has $\mathbb{E}[X]=0$ and $\text{Var}[X]=1$, with the distribution remaining unchanged (i.e., $X$ is still the uniform distribution). As such, the standardized uniform distribution $x \sim \mathcal{U}[-\sqrt{12}/2,\sqrt{12}/2]$, resulting in $h(X) = \log \sqrt{12}$. Using Eqn. \eqref{eqn:SLB}, the final resulting Shannon lower bound for the case $d_x=1$ is defined as:

\begin{equation}
\label{eqn:SLB_final}
R_{SLB}(D) = \max \left\{0, \frac{1}{2} \log \frac{6}{\pi eD} \right\} \, .
\end{equation}

The same steps can be followed to derive the rate-distortion theory for general $d_x$ cases \citep{gray1989source}, where the result will be identical to Eqn. \eqref{eqn:SLB_final} if the distortion measure is defined as Eqn. \eqref{eqn:distortion}. 

Note we have derived eqn. \eqref{eqn:SLB} final in the context of continuous uniform sources with dimensions $d_x$. In particular, notice that $R_{SLB}(D) \to \infty$ as $D \to \infty$. Intuitively, you need infinite nats (or bits in $\log_2$) to represent any real number exactly. As such, a close to zero reconstruction as seen is Fig. \ref{fig:rate_distortion}b is due to finite numerical precision.

\begin{table}[h]
\centering
\begin{tabular}{ll}
\hline \\ 
\textbf{Hyperparameter} & \textbf{Values} \\ \\ \hline \\
Hidden layer sizes & $(100)$, $(100,100)$, $(100,100,100)$, $(100,100,100,100)$ \\ 
Activation & ReLU \\
Solver & Adam \\
Tolerance (tol) & $10^{-6},\,10^{-8},\,10^{-10}$ \\
L2 regularization ($\alpha$) & $10^{-4},\,10^{-3},\,10^{-2},\,10^{-1}$ \\
Learning rate schedule & Adaptive \\ \\ \hline \\
\end{tabular}
\caption{Neural network hyperparameter grid.}
\label{tab:hyperparams}
\end{table} 

\section{Hyperparameters for Neural Network Decoder}
\label{appendix:neural_network}
In this work we use a simple multilayer perceptron to map from sensor values $\bm{Y}$ to reconstructed load $\bm{X}$. We use sklearn's \citep{scikit-learn} \verb|MLPRegressor| to fit the neural networks to the data. Sklearn's \verb|GridSearchCV| is applied to obtain the best set of hyperparameters for each information channel. The set of available hyperparameters is presented in Table \ref{tab:hyperparams}. All the neural networks are trained using the Adam optimizer \citep{Kingma2014AdamAM} to $5000$ epochs, with early stopping employed to avoid overfitting. In this work, we use $4000$ data points to train the data with $5$-fold cross validation and $1000$ data points as test data used to present Fig. \ref{fig:rate_distortion}.

\section{Finite Element Analysis}
\label{appendix:fea}
In this Section, we briefly outline the finite element simulations used in this work. We implemented our finite element simulations with FEniCSx \citep{AlnaesEtal2014, baratta2023dolfinx, BasixJoss}, and generated all meshes with Gmsh \citep{geuzaine2009gmsh}. The elastic bodies have length $L=100$ and width $W=100$. The constitutive model used is a simple linear elastic material model with Young's modulus $E=100$ and Poisson's ratio $\nu = 0.0$. Analogous to the elastic halfspace example, loads with magnitude $F=1$ and width $a=100$ (i.e., the width of the entire top boundary) are applied on the top of the structure. The bottom of the structure is fixed in both the $\hat{e}_1$ and $\hat{e}_2$ directions. We discretize each domain with a triangular meshes and quadratic basis functions. Since calculating mutual information requires a distribution of $\sigma_{22}$ at the sensor locations, to save computational time and cost, for each domain case, we factorize the corresponding linear system once with PETSc's Cholesky decomposition~\citep{balay2019petsc} and reuse the same matrix factorization for each different applied loading case (i.e., we reuse the factorized left-hand side of the linear system and only change the right-hand side of the linear system). 

\subsection{Mesh Refinement Study}
\label{appendix:mesh_refine}
\begin{figure}[ht]
    \centering
    \includegraphics[width= \textwidth]{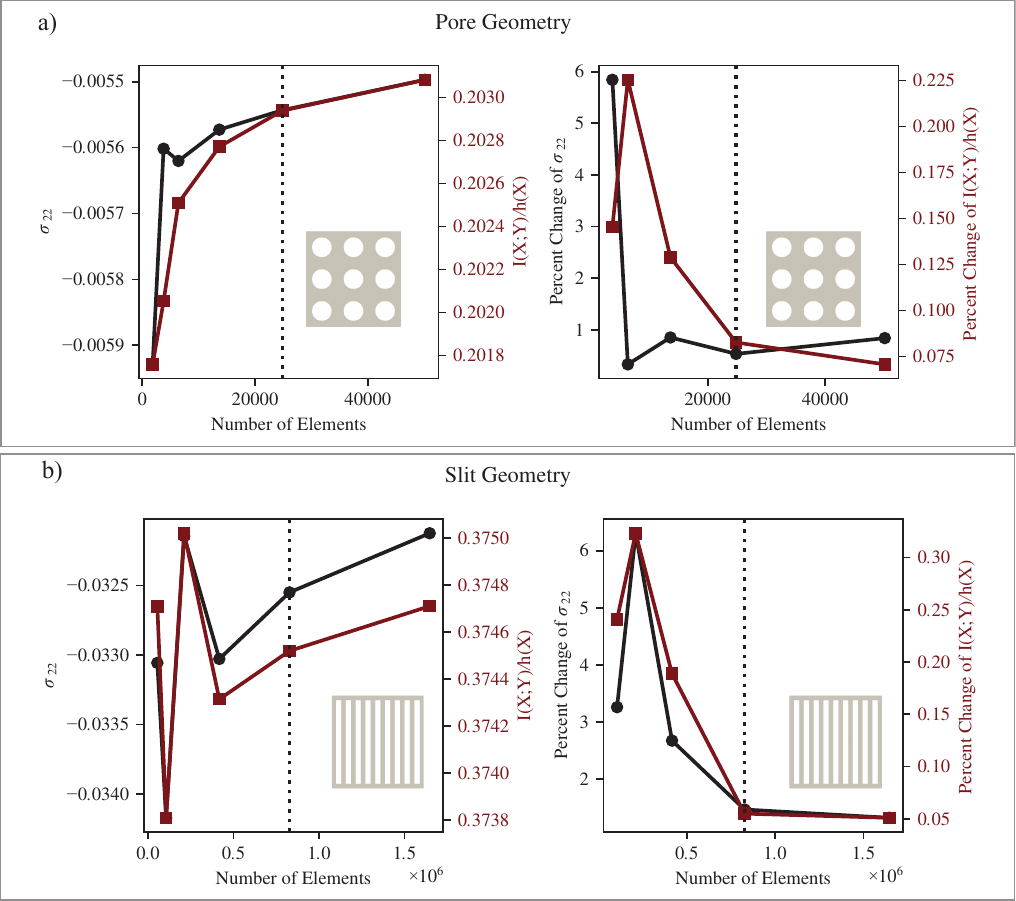}
    \caption{Results from our mesh refinement study. a) Mesh refinement study of pore geometry. Left panel plots $\sigma_{yy}$ (black) and $I(X;Y)/H(X)$ (gray) with mesh density. Right panel shows the percent change of $\sigma_{yy}$ (black) and $I(X;Y)/H(X)$ (gray) with mesh density. Dotted line shows selected mesh density. b) Mesh refinement study of slit geometry. Left panel plots $\sigma_{yy}$ (black) and $I(X;Y)/H(X)$ (gray) with mesh density. Right panel shows the percent change of $\sigma_{yy}$ (black) and $I(X;Y)/H(X)$ (gray) with mesh density. Dotted line shows selected mesh density.}
    \label{fig:mesh_refinement}
\end{figure}

To make sure that our FEA results are agnostic to mesh density, we perform mesh refinement studies on representative examples from the pore geometry class and slit geometry class. Specifically, we select a structure with $3$ units for pore geometry and $9$ units for slit geometry. Both structure types are subjected to loads with $d_x=6$ (i.e., loads constructed with $6$ term Legendre polynomials). The mesh density is controlled with the Gmsh parameter \verb|Mesh.CharacteristicLengthMin| and \verb|Mesh.CharacteristicLengthMax| to a characteristic length $l_c$ value. For the pore geometry, $l_c = L/(n_{units}l_{den})$ where $l_{den}$ is the parameter controlling the mesh density and $n_{num}$ is the number of units. For the slit geometry, $l_c = L/(20l_{den})$ since $L/20$ is the width of each pillar. This $l_{den}$ selection ensures that the mesh refinement studies will extend to different numbers of units for each geometry type. To perform the mesh refinement study, we approximately doubled the element density by incrementally increasing the characteristic length $l_{den}$ by a factor of $\sqrt{2}$. We both used the stress at point $(-L/2,0)$ (i.e., at one of the sensor locations) and the normalized mutual information $I(\bm{X};\bm{Y})/h(\bm{X})$ computed between the applied load $t$ and the sensor readings as our quantities of interest (QoI). To track the change of QoIs, we define the percent change in QoI as: 

\begin{equation}
\frac{\text{QoI}^{current} - \text{QoI}^{prev}}{\text{QoI}^{prev}} \times 100\%
\end{equation}
where $\text{QoI}^{prev}$ and $\text{QoI}^{current}$ are the previous and current values of the QoI respectively.

Fig. \ref{fig:mesh_refinement}a) and Fig. \ref{fig:mesh_refinement}b) plot our mesh refinement study for the pore geometry and slit geometry respectively. We concluded that for both geometries, $l_{den} = 40$ is sufficient for a converged solution of less than $1.5\%$ change in $\sigma_{22}$, while the normalized mutual information $I(\bm{X};\bm{Y})/h(\bm{X})$ has a much smaller percent change between mesh densities. 

\section{Principal Stress Lines}
\label{appendix:psl}
\begin{figure}[ht]
    \centering
    \includegraphics[width= \textwidth]{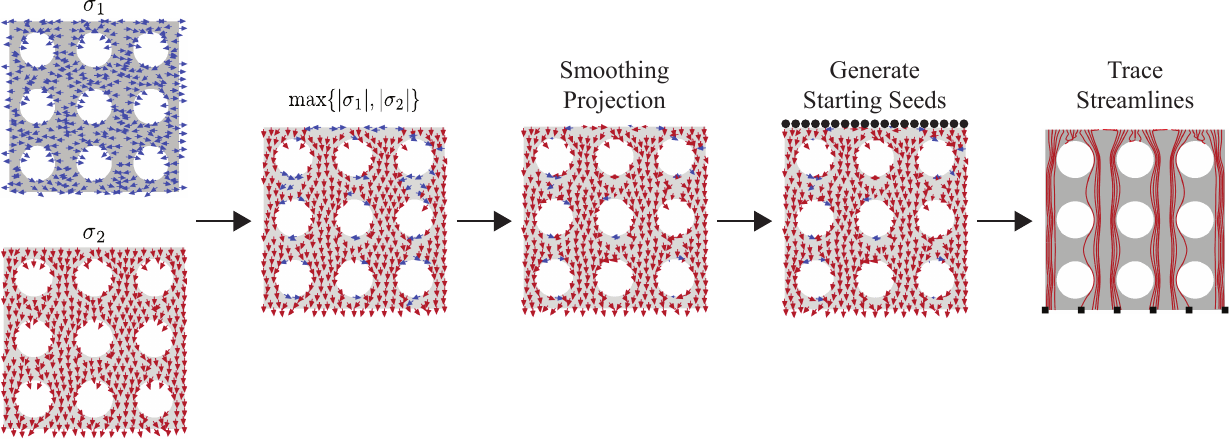}
    \caption{Pipeline to compute the principal stress lines used in this work. The major ($\sigma_1$) and minor ($\sigma_2$) principal axis are first computed for the structure under uniform load. The maximum magnitude between the major and minor principal stresses is then selected and smoothed into a continuous function space. To start tracing principal stress lines in the resulting vector field, we first put seeds evenly spaced on top of the structure and then performed numerical integration of the vector field to visualize the principal stress lines.}
    \label{fig:PSL}
\end{figure}

Principal stress lines are a simple method to visualize load paths and have been applied in many areas such as topology optimization \citep{kwok2016structural,yan2025principal}. However, obtaining principal stress is typically an ad-hoc process and the specific approach can vary slightly across application domains. Here we outline our approach to obtaining the principal stress lines.

We first compute the stress field under uniform compression. Uniform compression is considered because, for the sets of applied loads $\mathcal{X}_{\text{full}}$ and $\mathcal{X}_{\text{even}}$ with $d_x$ coefficients, the expected load satisfies $\mathbb{E}[X] = \mathbf{0}$, since $X \sim \mathcal{U}^{d_x}(-10, 10)$ (see Section~\ref{meth:load}). This leads to an effective uniform compressive load with magnitude $m$. Because this uniform load corresponds to the expected load distribution, linear superposition implies that the resulting uniform stress field is also the expected stress field for both sets of applied loads. Consequently, the principal stress lines generated under uniform compression represent the expected load paths for both load sets.

Following Fig \ref{fig:PSL}, we first obtain the principal stress and principal directions by post-processing the FEA results. More specifically, given a resultant 2D stress field from FEA, the principal stresses $\sigma_{1,2}$ (note that the comma here does not denote differentiation but denotes the major and minor principal stress $\sigma_1$ and $\sigma_2$ respectively) and directions $\theta_p$ can be obtained by~\citep{kwok2016structural}:

\begin{subequations}
\begin{align}
\sigma_{1,2} &= \frac{\sigma_{11} + \sigma_{22}}{2} \pm \sqrt{\frac{\sigma_{11} - \sigma_{22}}{2} + \sigma_{12}^2} \label{eqn:principal_stress} \\ 
\tan 2\theta_p & = \frac{2\sigma_{12}}{\sigma_{11}-\sigma_{22}} \label{eqn:principal_directions}
\end{align}
\end{subequations}

Crucially, while it is trivial to find which of the principal stresses corresponds to the major and minor principal stress from Eqn. \eqref{eqn:principal_stress}, the principal \emph{directions} is more convoluted since Eqn. \eqref{eqn:principal_directions} will have two feasible solutions $\theta_p$ and $\theta_p+\pi/2$, with each solution corresponding to a major or minor principal direction. However, it is not clear how to differentiate which directions belong to the major or minor directions. In this work, we determine the principal directions by looking at the magnitude of $\sigma_{11}$ and $\sigma_{22}$ where~\citep{yan2025principal}:

\begin{subequations}
\begin{align}
    \label{eqn:principal_directions1}
    \theta_{1} = 
    \begin{cases}
    \theta_p & \text{if} \quad \sigma_{11} > \sigma_{22} \\
    \theta_p+\frac{\pi}{2} & \text{if} \quad \sigma_{11} < \sigma_{22}
    \end{cases}
\end{align}
\begin{align}
\label{eqn:principal_directions2}
    \theta_{2} = 
    \begin{cases}
    \theta_p & \text{if} \quad \sigma_{11} < \sigma_{22} \\
    \theta_p + \frac{\pi}{2} & \text{if} \quad \sigma_{11} > \sigma_{22} 
    \end{cases}
\end{align}
\end{subequations}

where $\theta_{1}$ and $\theta_2$ correspond to the major and minor principal directions respectively. In this work, we are interest in how loads propagate from the location of application to the boundary conditions. As such, rather than tracing the major and minor principal stresses individually which is popular in topology optimization \citep{kwok2016structural, yan2025principal}, we instead combine the two principal directions fields by selecting the maximum magnitude between the principal directions such that the final vector field of principal directions is denoted by: 

\begin{equation}
\theta_{final} = 
\begin{cases}
\theta_1 & \text{if} \quad |\sigma_1| \geq |\sigma_2| \\
\theta_2 & \text{if} \quad |\sigma_1| \leq |\sigma_2|
\end{cases}
\end{equation}

Note that the vector field $\theta_{final}$ is currently in DG1 (i.e., discontinuous linear elements) space following the stress solution, since we ran our FEA simulations with CG2 (i.e., the displacements are in quadratic Lagrange elements). Consequently, the current vector field in not suitable to trace principal stress lines since principal stress lines rely on integrating over smooth vector fields \citep{kwok2016structural, yan2025principal}. To smooth the vector field, we use $\ell_2$ Galerkin projection to map the discontinuous vector field into a continuous field~\citep{hughes2003finite}, which results in solving a simple mass-matrix system in FEniCSx. Once the field is relatively smooth, we put initial seed sources evenly-spaced at the point of load application and interpret the resulting vector field as a velocity field. The velocity field is integrated with the fourth-order Runge-Kutta (i.e., RK4) to finally trace the principal stress lines. The integration is done using PyVista~\citep{sullivan2019pyvista}, an open-sourced python-based interface for Visualization Toolkit (VTK). Specific integration parameters and functions used can be found on our Github Page \url{https://github.com/pprachas/info_mech}. 

\section{Additional Details on Bayesian optimization}
\label{appendix:opt}
\begin{figure}[ht]
    \centering
    \includegraphics[width= \textwidth]{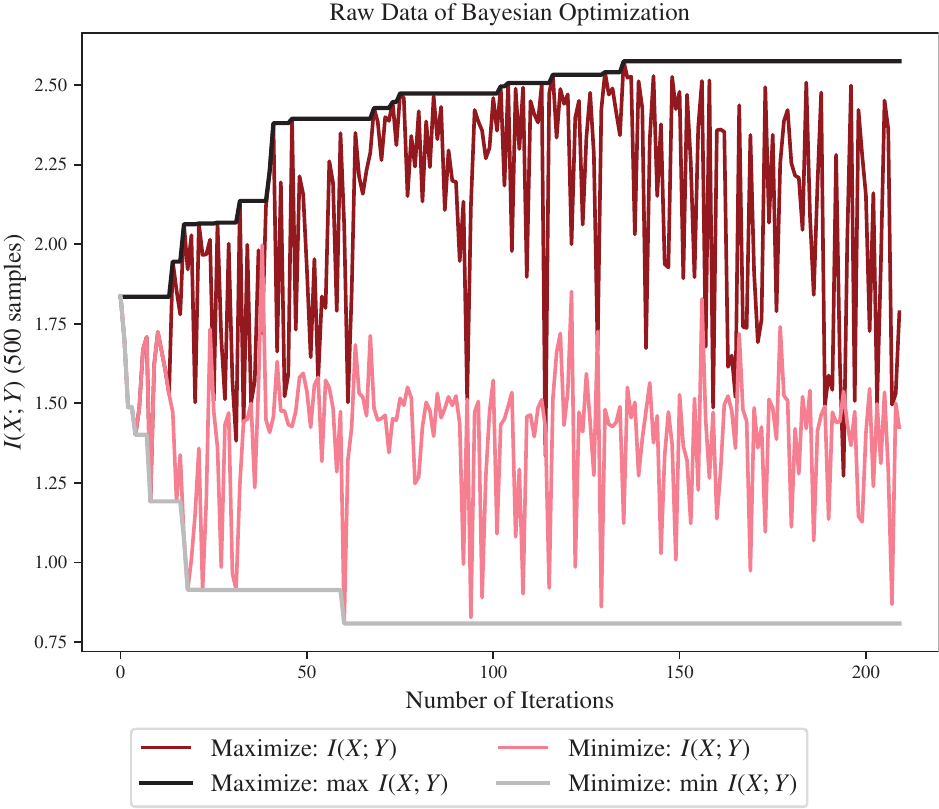}
    \caption{Raw data from Bayesian optimization. The black line plots data from the maximization optimizer, while the gray line plots data from the minimization optimizer. The red and pink lines show the actual $I(X;Y)$ computed at each iteration for the maximization and minimization problem respectably, while the black and gray lines shows the cumulative results for the maximization and minimization problem respectably.}
    \label{fig:raw_opt}
\end{figure}

Here, we summarize the parameters that we used for Bayesian Optimization in Section \ref{result:optimize}. Bayesian optimization is a data-driven optimization approach that is designed for black-box derivative free global optimization \citep{brochu2010tutorial,frazier2018tutorial}. In brief, Bayesian optimization identifies optimal design parameters through an iterative process that selects promising candidate designs and updates a probabilistic surrogate model of the objective. More formally, Bayesian Optimization seeks solve the optimization problem where:

\begin{equation}
z^\star = \max_{z \in \mathcal{Z}} f(z) \, , 
\end{equation}

where $z \in \mathbb{R}^d$ is the design variables, $z^{\star}$ is the optimal point, $\mathcal{Z}$ is the set of feasible designs, and $f$ is the objective function \citep{brochu2010tutorial, frazier2018tutorial}. Bayesian optimization, due to the curse of dimensionality, is most effective in low to moderate dimensions i.e., $d\leq 20$, and $\mathcal{Z}$ is typically a hyper-rectangle ${z \in \mathbb{R}^d: p_i \leq z_i \leq q_i}$ \citep{frazier2018tutorial}. The objective function $f$ is also ideally continuous, since we will be representing $f$ with a probabilistic surrogate. In this work, we use Gaussian Processes (GPs) as our surrogate model such that the target $f(z) \sim \mathcal{N}(\mu(z),\sigma^2(z))$ (i.e., target $f(z)$ has a normal distribution with mean $\mu(z)$ and variance $\sigma^2(z)$). with a Mat\'ern kernel \citep{williams2006gaussian}.

\begin{equation}
k(z, z') =
\theta_0 \frac{2^{1-\nu}}{\Gamma(\nu)}
\left(\frac{\sqrt{2\nu}}{l}\,\lVert z - z' \rVert\right)^{\nu}
K_{\nu}\!\left(\frac{\sqrt{2\nu}}{l}\,\lVert z - z' \rVert\right) \, ,
\end{equation}

where $k(z,z')$ is the kernel function, $K_\nu$ is the modified Bessel function, $l$ is the length-scale parameter optimized during the fitting process, and $\theta_0$ are the kernel parameters that are fit to the data \citep{frazier2018tutorial}. In this work, we set $\nu = 2.5$ to get the Mat\'ern $2.5$ kernel, a common kernel for GPs in Bayesian optimization \citep{brochu2010tutorial}. 

In this work, the objective is to maximize or minimize mutual information. Estimating mutual information requires a large number of samples from the finite-element solver; accordingly, we use $N=500$ samples to approximate the mutual information and treat the resulting estimate as noisy. This uncertainty is incorporated into the Bayesian optimization procedure as an observation noise variance in sklearn’s \verb|GaussianProcessRegressor|. The variance of the mutual information estimator scales as $O(1/N)$ according to theoretical results \citep{gao2015efficient}, and we use this scaling to set the noise level. Although this variance estimate is formally derived for the KSG estimator with the $\ell_2$ distance metric, whereas we follow the original KSG formulation and use $\ell_\infty$ (see Appendix \ref{appendix:ksg}), we find the theoretical noise estimate sufficient for the purposes of Bayesian Optimization in this work.

The second main component of Bayesian Optimization is the acquisition function. The acquisition function is responsible for deciding where to sample next. While there are many choices of acquisition function, in this work, we use the Expected Improvement (EI) as our acquisition function. The acquisition function for EI is:

\begin{equation}
\alpha = \mathbb{E}[\max \{ 0, f(z)-f_{max}\}] \, ,
\end{equation}

where $f_{max}$ is the largest value \emph{observed so far} \citep{brochu2010tutorial,frazier2018tutorial}. Intuitively, The acquisition function $\alpha$ aims to pick the point that will give the biggest expected improvement from the current optimum. Since target $f(z)$ is approximated by a GP, EI can be rewritten as:

\begin{equation}
\mathrm{EI}(z; \xi)
= (\mu(z) - f_{max} - \xi)\, \Phi\!\left(\frac{\mu(z) - f_{max} - \xi}{\sigma(z)}\right)+ \sigma\, \phi\!\left(\frac{\mu(z) - f_{max} - \xi}{\sigma(z)}\right) \, ,
\end{equation}

where $\Phi$ and $\phi$ are the cumulative distribution function and probability distribution function of a standard normal distribution respectively, and $\xi$ is a prescribed parameter \citep{brochu2010tutorial, frazier2018tutorial}. Intuitively, the first term favors picking points near the expected maximum (i.e., exploitation), while the second term favors picking points where the surrogate has a high variance (i.e., exploration). As such, $\xi$ is a hyperparameter that control the exploration vs. exploitation tradeoff, with a larger value of $\xi$ favoring exploration. In this work, we set $\xi = 0.01$ which is a good starting point for most Bayesian Optimization problems \citep{brochu2010tutorial}.

The void fraction constraints in Section \ref{result:optimize} are implemented with \verb|Bayesian Optimization|'s \citep{bayesopt} implementation of Bayesian Optimization with inequality constraints \citep{gardner2014bayesian}. Note that we only have to implement the lower, as the upper bound is already constrained by the bounds we give the design variables. 

\changes{\section{The Inextensible Elastica}
\label{appendix:elastica}}
\changes{In this Section, we review the governing equations of the inextensible elastica, and outline our numerical scheme to solve the corresponding nonlinear equations. The shape of an elastica with length $L$ is described through a parameterized curve $s \in [0,L]$~\citep{bigoni2015new, timoshenko2012theory}. Following Fig \ref{fig:elastica}a, the elastica is loaded with axial force $F_1$ , shear force $F_2$, and an applied moment $M$ at $s=L$ while being clamped at $s=0$. The governing equation of a planar elastica is expressed as:}
\changes{
    \begin{equation}
    \theta''(s) - \frac{F_1}{EI}\sin\theta(s)
    + \frac{F_2}{EI}\cos\theta(s) = 0 \, .
\end{equation}}

\changes{The centerline coordinates $u$ and $v$ are related to the rotation field by:}
\changes{
\begin{subequations}
\begin{align}
u'(s) &= \cos\theta(s) \\
v'(s) &= \sin\theta(s) \, .
\end{align}
\end{subequations}}

\changes{The corresponding boundary conditions are:}

\changes{
\begin{subequations}
\begin{align}
u(0) &= 0, \\
v(0) &= 0, \\
\theta(0) &= 0, \\
EI\,\theta'(L) &= M \, ,
\end{align}
\end{subequations}}

\changes{For simplicity, we set $EI=1$ and $L=1$. Note that the range of $F_1$ is chosen to avoid buckling bifurcations by making sure the magnitude of the compressive load is below the Euler critical buckling load $P_{cr}$ of a fixed-free beam where:}
\changes{
\begin{equation}
P_{cr} = \frac{\pi^2EI}{(2L)^2}\approx 2.47 \, .
\end{equation}}
\changes{We solve the governing equations with \texttt{scipy.integrate.solve\_bvp}~\citep{virtanen2020scipy} and extract $\theta(s)$, $u(s)$, and $v(s)$. To aid in convergence without numerical continuation, we initialize the solution with the analytical solution derived from superposition of Euler-Bernoulli beams (i.e., using $M=EI\kappa$) with the same boundary conditions outlined in Fig \ref{fig:elastica}a where:}

\changes{\begin{subequations}
\begin{align}
    \theta_{EB}(s) &= \frac{1}{EI}\!\left[M\,s + F_2\!\left(L\,s - \tfrac{s^2}{2}\right)\right], \label{eq:eb-theta} \\
    u_{EB}(s) &= s, \label{eq:eb-x} \\
    v_{EB}(s) &= \frac{1}{EI}\!\left[\tfrac{M\,s^2}{2} + F_2\!\left(\tfrac{L\,s^2}{2} - \tfrac{s^3}{6}\right)\right]. \label{eq:eb-y}
\end{align}
\end{subequations}}
\changes{Note that the axial force $F_1$ does not enter the Euler-Bernoulli equations.}

\bibliographystyle{plain}
\bibliography{main}
\end{document}